\documentclass[twocolumn,preprintnumbers,amsmath,amssymb,showkeys,mathtools,letterpaper]{revtex4-1}

\usepackage{graphicx}
\usepackage{upgreek}
\usepackage{amsmath}
\usepackage{amssymb}
\usepackage{dcolumn}
\usepackage{chngpage}

\usepackage{graphicx}
\usepackage{dcolumn}
\usepackage{bm}
\usepackage{amssymb} 
\usepackage{subfigure}
\usepackage{color}
\usepackage{epstopdf}
\usepackage{tabularx}
\newcolumntype{C}{>{\centering\arraybackslash}X}
\usepackage{array,booktabs}

\begin{document}

\preprint{Preprint}

\title{Molecular Dynamics Simulations of NMR Relaxation and Diffusion of Bulk Hydrocarbons and Water}
\author{P. M. Singer}\email{ps41@rice.edu}
\author{D. Asthagiri}\email{dna6@rice.edu}
\author{W. G. Chapman}\email{wgchap@rice.edu}
\author{G. J. Hirasaki}\email{gjh@rice.edu}
\affiliation{Rice University, Department of Chemical and Biomolecular Engineering, 6100 Main St., Houston, TX 77005, USA}

\date{\today}


\begin{abstract}
Molecular dynamics (MD) simulations are used to investigate $^1$H nuclear magnetic resonance (NMR) relaxation and diffusion of bulk $n$-C$_5$H$_{12}$ to $n$-C$_{17}$H$_{36}$ hydrocarbons and bulk water. The MD simulations of the $^1$H NMR relaxation times $T_{1,2}$ in the fast motion regime where $T_1 = T_2$ agree with measured (de-oxygenated) $T_2$ data at ambient conditions, without any adjustable parameters in the interpretation of the simulation data. Likewise, the translational diffusion $D_T$ coefficients calculated using simulation configurations are well-correlated  with measured diffusion data at ambient conditions. The agreement between the predicted and experimentally measured NMR relaxation times and diffusion coefficient also validate the forcefields used in the simulation. The molecular simulations naturally separate intramolecular from intermolecular dipole-dipole interactions helping bring new insight into the two NMR relaxation mechanisms as a function of molecular chain-length (i.e. carbon number). Comparison of the MD simulation results of the two relaxation mechanisms with traditional hard-sphere models used in interpreting NMR data reveals important limitations in the latter. With increasing chain length, there is substantial deviation in the molecular size inferred on the basis of the radius of gyration from simulation and the fitted hard-sphere radii required to rationalize the relaxation times. This deviation is 
characteristic of the local nature of the NMR measurement, one that is well-captured by molecular simulations. 

\end{abstract}


\maketitle

\section{Introduction}\label{sc:Intro}

Molecular dynamics (MD) simulations of macromolecules such as proteins and polymers have become increasingly common as forcefields have improved and computers have become more powerful. Of particular interest is the integration of MD simulations with nuclear Overhauser effect spectroscopy (NOESY), and variants thereof, which directly probe the nuclear dipole-dipole interactions, and which have proven to yield unique information about the structure and dynamics of macromolecules \cite{bruschweiler:pnmrs1994,peter:jbnmr2001,luginbuhl:pnmrs2002,case:acr2002,kowalewski:book}. An important approach in the interpretation of non-rigid macromolecules has been the separation of the fast internal dynamics which are accessible by MD simulations, from the much slower dynamics related to rotation of the entire macromolecule. A popular technique for this separation has been the model-free Lipari-Szabo approach \cite{lipari:jacs1982}, which yields important information about the structure and dynamics of the non-rigid branches and internal motions of the macromolecules.

The integration of MD simulations and NMR measurements presented here also requires state of the art in both computation and measurement. On the measurement side, such studies require careful de-oxygenation of the fluid for accurate measurements of the intrinsic NMR relaxation times \cite{lo:SPE2002,shikhov:amr2016}, as well as diffusion experiments free of convection and hardware artifacts \cite{tofts:mrm2000,mitchell:pnmrs2014}. On the computational side, with current empirical potential (forcefield) models it is now possible to readily access $O(10-100)$~ns time scales that are of interest in understanding relaxation processes in confined fluids. But care is needed in ensuring the potential models are reasonable and the simulations are well-crafted. 

One emerging area of interest for integrating MD simulations with NMR measurement is in porous organic media such as kerogen, where the phase-behavior and transport properties of hydrocarbons and water are greatly affected by confinement in the organic nanometer-sized pores. Studies of fluid confinement in organic nano-pores is of great interest in characterizing organic-shale reservoirs such as gas-shale and tight-oil shale \cite{bui:spe2016}. For example, MD simulations are well suited for investigating the effects of wettability alterations in organic matter \cite{hu:spe2014}, or the effects of restricted diffusion and pore connectivity of organic nano-pores.
Recent MD simulations of fluids confined in graphitic nano-pores have shown promising results for molecular interactions between surfaces and fluids such as hydrocarbons and water \cite{hu:spe2014}. Recent NMR measurements of hydrocarbons and water confined in the organic nano-pores of isolated kerogen indicate that $^1$H--$^1$H dipole-dipole interactions play a key role in  surface relaxivity, pore-size analysis, and fluid typing \cite{singer:petro2016}. However, to the best of the authors' knowledge, no comprehensive studies have been performed which integrate MD simulations with NMR measurements of hydrocarbons and water, either in the bulk phase or under confinement in porous media, even though the techniques exist to generate molecular models of kerogen for MD simulations \cite{Ungerer:2015}.

Another emerging area of interest is in using MD simulations to characterize the NMR relaxation in complex multi-component fluids such as heavy crude-oils \cite{yang:jmr2008,yang:petro2012,chen:cpc2014,ordikhani:ef2016}, or model systems for crude-oils such as polymer-solvent mixes \cite{tutunjian:la1992}. In such cases the Lipari-Szabo model is a good approach for separating the slow dynamics related to viscosity from the fast dynamics related to local non-rigid molecular motions. The objective in such cases is to obtain a more fundamental understanding of the NMR relaxation and diffusion of crude-oils, thereby helping to characterize petroleum fluids.

The first step in embarking in such studies is to successfully integrate MD simulations and NMR measurements of relaxation and diffusion for bulk hydrocarbons and bulk water. This report presents such a task, and validates the forcefields used in MD simulations for bulk hydrocarbons and water, without any adjustable parameters in the interpretation of the simulation data. The MD simulations are then used to distinguish intramolecular from intermolecular NMR relaxation. Our studies provide new insight into the two relaxation mechanisms, which is not straightforward to do from NMR measurements alone \cite{woessner:jcp1964}. 
A comprehensive investigation into the two relaxation mechanisms is required to improve the characterization of macromolecules such as proteins and polymers, where typically only the intramolecular interaction is considered \cite{kowalewski:book}, without any substantial justification.

The rest of the article is organized as follows. The methodology behind the MD simulations of NMR relaxation and diffusion are presented in Section \ref{sc:Method}. This is followed by results and discussions in Section \ref{sc:Results}, including correlation between simulations and measurements, intramolecular versus intermolecular interactions, and, analysis using traditional hard-sphere models. Conclusions are presented in Section \ref{sc:Conc}.

\section{Methodology} \label{sc:Method}

This section presents details of the molecular simulation procedure, the NMR relaxation analysis, and the diffusion analysis. 

\subsection{Molecular Simulation}

The molecular simulations were performed using NAMD \cite{namd} version 2.11. The bulk alkanes were modeled using the CHARMM General Force field (CGenFF) \cite{cgenff} and bulk water was modeled using the TIP4P/2005 model \cite{tip4p}. In the case of water, the relaxation behavior (see below) using the more refined AMOEBA polarizable model \cite{amoeba14} was in excellent agreement with predictions based on TIP4P/2005 (data not shown). Given the typically greater challenge in modeling water, we thus expect the relaxation behavior in alkanes modeled with a polarizable model and the modern CGenFF non-polarizable model to be comparable. Thus for computational economy, in this study we used non-polarizable models throughout. (The good agreement between simulations and experiments supports this suggestion.) 

The simulation cells for the hydrocarbons were constructed as follows. We first construct a single molecule using the Avogadro program \cite{avogadro}. We than create $N$ copies of this molecule and pack it into a cube of volume $V$ using the Packmol program \cite{packmol}. The volume is chosen such that the number density $N/V$ corresponds to the experimentally determined number density at 293.15~K (20$^{\rm o}$C). All the subsequent simulations are performed at this number density and temperature due to the availability of density \cite{nist,*crc} and viscosity \cite{viscosity} data at that temperature. This initial system was energy minimized to relieve any energetic strain due to limitations in the packing procedure. For the water simulation, we used an existing equilibrated simulation cell for the present study.  The alkane simulations typically 
comprised between 1660 ($n$-C$_5$H$_{12}$) and 2125 ($n$-C$_{17}$H$_{36}$) carbon atoms, while the water system comprised 
512 water molecules. 

For the molecular dynamics simulations, the equations of motion were propagated using the Verlet algorithm with a time step of 1.0~fs. (The SHAKE \cite{md:shake} procedure was used to constrain the geometry of water molecules, but the C--H bond in the alkane was not similarly constrained.) 
We first equilibrate the simulation cell for 2~ns at a temperature of 293.15~K.  The temperature was controlled by reassigning velocities (obtained from a Maxwell-Boltzmann distribution) every 250~fs. The Lennard-Jones interactions were terminated at 14.00~{\AA} (11~{\AA} for water) by smoothly switching to zero starting at 13.00~{\AA} (10~{\AA} for water). Electrostatic interactions were treated with the particle mesh Ewald method with a grid spacing of  0.5 {\AA}; the real-space contributions to the electrostatic interaction were cutoff at 14.00~{\AA}. For water the equilibration was for 0.25~ns.

We started the production run of 2~ns time from the end-point of the equilibration run. Since the dynamics of relaxation are of first interest here, 
the production phase was performed at constant density and energy (NVE ensemble), i.e.\ the system was not thermostated. The average temperature of the system is within 2~K of the target (293.15~K) for all the alkanes except $n$-C$_{5}$H$_{12}$ (pentane) for which the deviation was about 3~K.  As a further check of the simulation,  we also monitored the average energy of the system. All the simulations showed excellent energy conservation, with the standard error of the mean energy relative to the mean energy being less than $5\times 10^{-6}$.  In the production phase, configurations were saved every 0.1~ps for the hydrocarbons and 0.125~ps for water.

\subsection{NMR Relaxation}\label{sc:Back}

\begin{figure}[h!]
	\begin{center}
		\includegraphics[width=0.9\columnwidth]{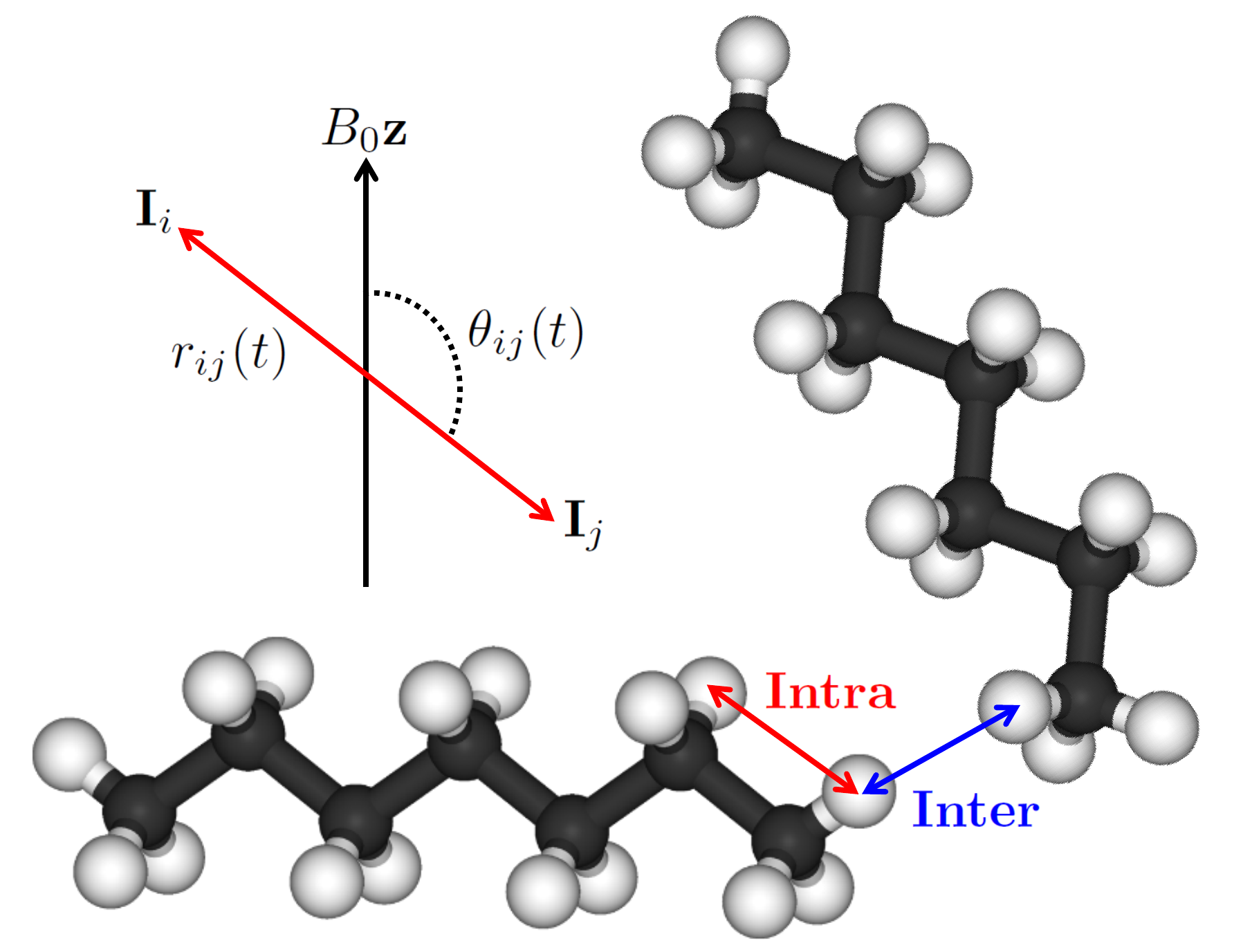} 
	\end{center}
	\caption{Illustration of $n$-C$_{7}$H$_{16}$ (heptane) molecules, with hydrogen (white) and carbon (black) atoms. A $^1$H nuclear spin is shown interacting with a neighboring $^1$H nuclear spin on the same heptane molecule, known as the intramolecular dipole-dipole interaction (red) which is characterized by the rotation correlation time $\tau_R$ of a single molecule. The same $^1$H nuclear spin is also shown interacting with a $^1$H nuclear spin from a different heptane molecule, known as the intermolecular dipole-dipole interaction (blue) which is characterized by the translational correlation time $\tau_T$ between molecules. Axes illustrate the distance $r_{ij}(t)$ between nuclear spins {\bf I$_i$} and {\bf I$_j$} with respect to the applied static magnetic-field direction ${\bf B}_0 = B_0 \bf z$, and corresponding polar angle $\theta_{ij}(t)$. }
	\label{fg:Hept}
\end{figure}

The theory of NMR relaxation in liquids is very well established \cite{bloembergen:pr1948,torrey:pr1953,abragam:book,mcconnell:book,cowan:book,kimmich:book}, and for brevity we consider only the essential details. 
The autocorrelation function $G^m(t)$ of fluctuating magnetic dipole-dipole interactions is central to the development of NMR relaxation theory in liquids. This autocorrelation is also well suited for computation using MD simulations \cite{peter:jbnmr2001,case:acr2002}. Using the convention 
in the text by McConnell \cite{mcconnell:book}, $G_{R,T}^m(t)$ (in units of s$^{-2}$) is given by
\begin{multline}
G_{R,T}^m(t) = \frac{3\pi}{5} \! \left(\frac{\mu_0}{4\pi}\right)^2 \! \hbar^2 \gamma^4  \frac{1}{N_{R,T}}\sum\limits_{i \neq j}^{N_{R,T}}  \\  \left< \frac{Y_2^m({\bf \Omega}_{ij}\!\left(t+\tau\right))}{r_{ij}^3\!\left(t+\tau\right)}  \frac{Y_2^{m*}({\bf \Omega}_{ij}\!(\tau))}{r_{ij}^3\!(\tau)} \right>_{\!\! \tau},
\label{eq:GmRTY}
\end{multline}
where $t$ is the lag time in the autocorrelation, $\mu_0$ is the vacuum permeability, $\hbar$ is the reduced Planck constant, and $\gamma$ is the
gyromagnetic ratio. ${\bf \Omega} \equiv (\theta, \phi)$, where $\theta$ is the polar angle (see Fig.~\ref{fg:Hept}) and $\phi$ the azimuthal angle. The ensemble average is performed by a double summation over spin pairs $i$ and $j$ of the nuclear ensemble, with $i\neq j$. The subscripts $R$ and $T$ correspond to the partial ensembles $N_R$ and $N_T$ for intramolecular or intermolecular $^1$H--$^1$H dipole-dipole interactions, respectively. This is illustrated in Fig. \ref{fg:Hept}, 
where the intramolecular interaction corresponds to $N_R = 16$ in the case of heptane ($n$-C$_{7}$H$_{16}$), while $N_T$ corresponds to $^1$H in all other molecules. The partition into partial ensembles $N_R$ and $N_T$ is a straight-forward book-keeping step in MD simulations, and leads to a clear separation of $G_{R}(t)$ and $G_{T}(t)$. The brackets in Eq. \ref{eq:GmRTY} refer to the average over recorded time $\tau$ during the production run:
\begin{equation}
\left< . . . .^{{\strut}{}} \right>_{\!\! \tau} = \frac{1}{N_{\tau}}\sum_{\tau = 0}^{N_{\tau}} . . . 
\label{eq:Ergodic}
\end{equation}
where $\tau$ is recorded in 0.1 ps increments, with a total of $N_{\tau} = $20000 for hydrocarbons and $N_{\tau}=16000$ for water.

For spherically symmetric systems such as the bulk fluids in question, the autocorrelation function $G_{R,T}^m(t) = G_{R,T}(t)$ is independent of the order $m$, which amounts to saying that the direction of the applied magnetic field ${\bf B}_0 = B_0 \bf z$ is arbitrary. 
For simplicity, we study the $m = 0$ harmonic $Y_2^0({\bf \Omega}) = \sqrt{5/16\pi}\, (3\cos^{2}\theta-1) $ \cite{mcconnell:book} 
by MD simulations. Thus the final expression for the autocorrelation used here is: 
\begin{multline}
G_{R,T}(t) = \frac{3}{16} \! \left(\frac{\mu_0}{4\pi}\right)^2 \! \hbar^2 \gamma^4  \frac{1}{N_{R,T}}\sum\limits_{i \neq j}^{N_{R,T}}  \\  \left< \frac{(3\cos^{2}\!\theta_{ij}\!(t+\tau)-1)}{r_{ij}^3\!\left(t+\tau\right)}  \frac{(3\cos^{2}\!\theta_{ij}\!(\tau)-1)}{r_{ij}^3\!(\tau)} \right>_{\!\! \tau}.
\label{eq:GmRT}
\end{multline}

From the stored simulation trajectory, both an ensemble average over $N_R$ or $N_T$ and a time average over $\tau$ were performed in Eq. \ref{eq:GmRT}. To economize on computational time used in the analysis, for all the
systems, only a subset of molecules was sampled for constructing the above average. For a chosen molecule, a hydrogen is first selected and its separation ($r_{ij}$) from all other hydrogen nuclei on the chosen molecule is recorded (intramolecular), as well as the separation from all other hydrogen nuclei in the rest of the fluid (intermolecular). Such a record is then constructed for all time points $\tau$. From
the record of $r_{ij}(\tau)$ and $\cos \theta_{ij}(\tau)$ (Fig.~\ref{fg:Hept}), 
we calculated $G_{R,T}(t)$ at different lag times $t$ using fast Fourier transforms for computational efficiency (the code is available upon request). 
The computed lag time ranged from 0 ps $ \leq t \leq $ 150 ps in increments of 0.1~ps for hydrocarbons and 0 ps $ \leq t \leq $ 50 ps for water in increments of 0.125 ps.

\begin{figure}[h!]
	\begin{center}
		\includegraphics[width=0.9\columnwidth]{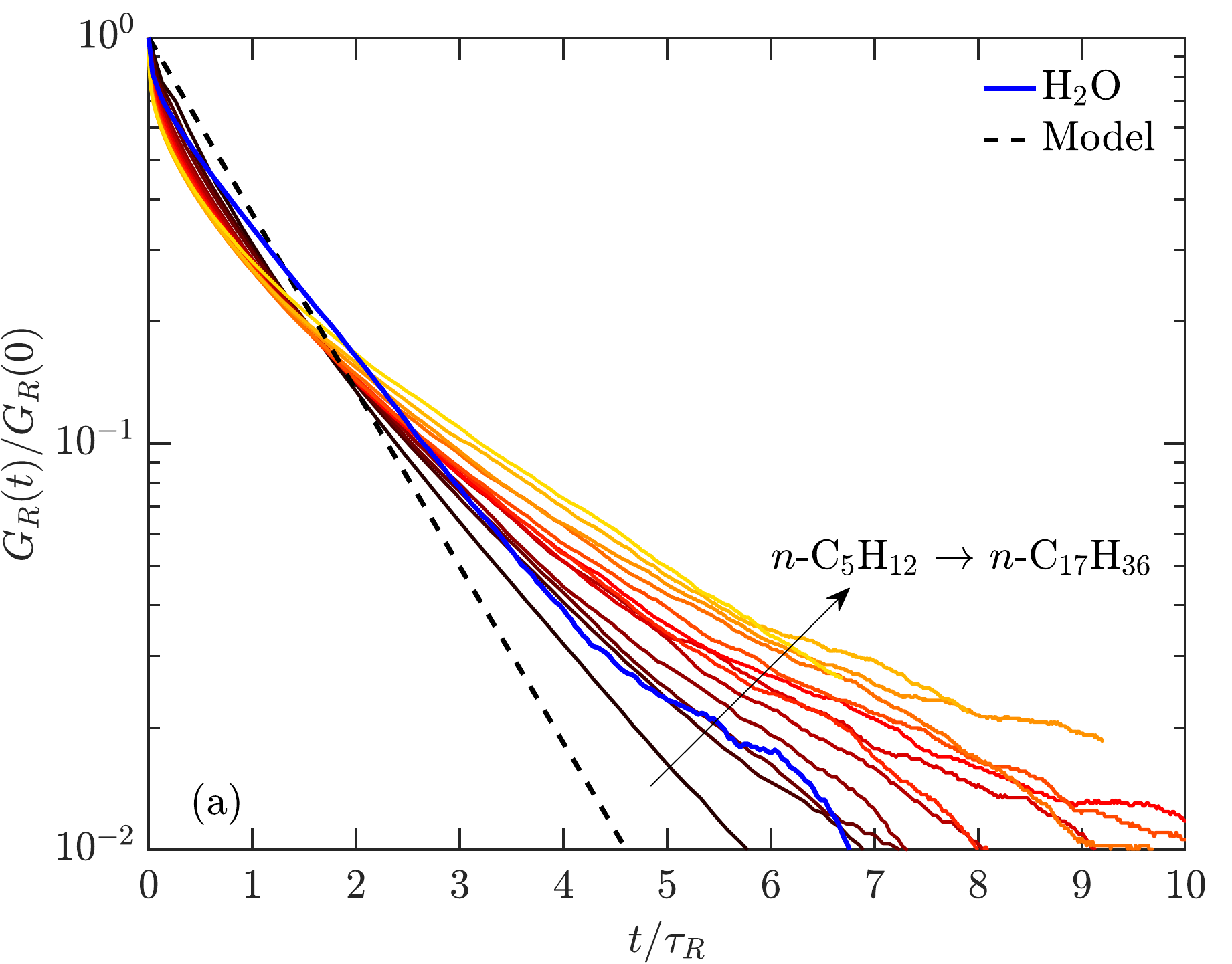} 
		\includegraphics[width=0.9\columnwidth]{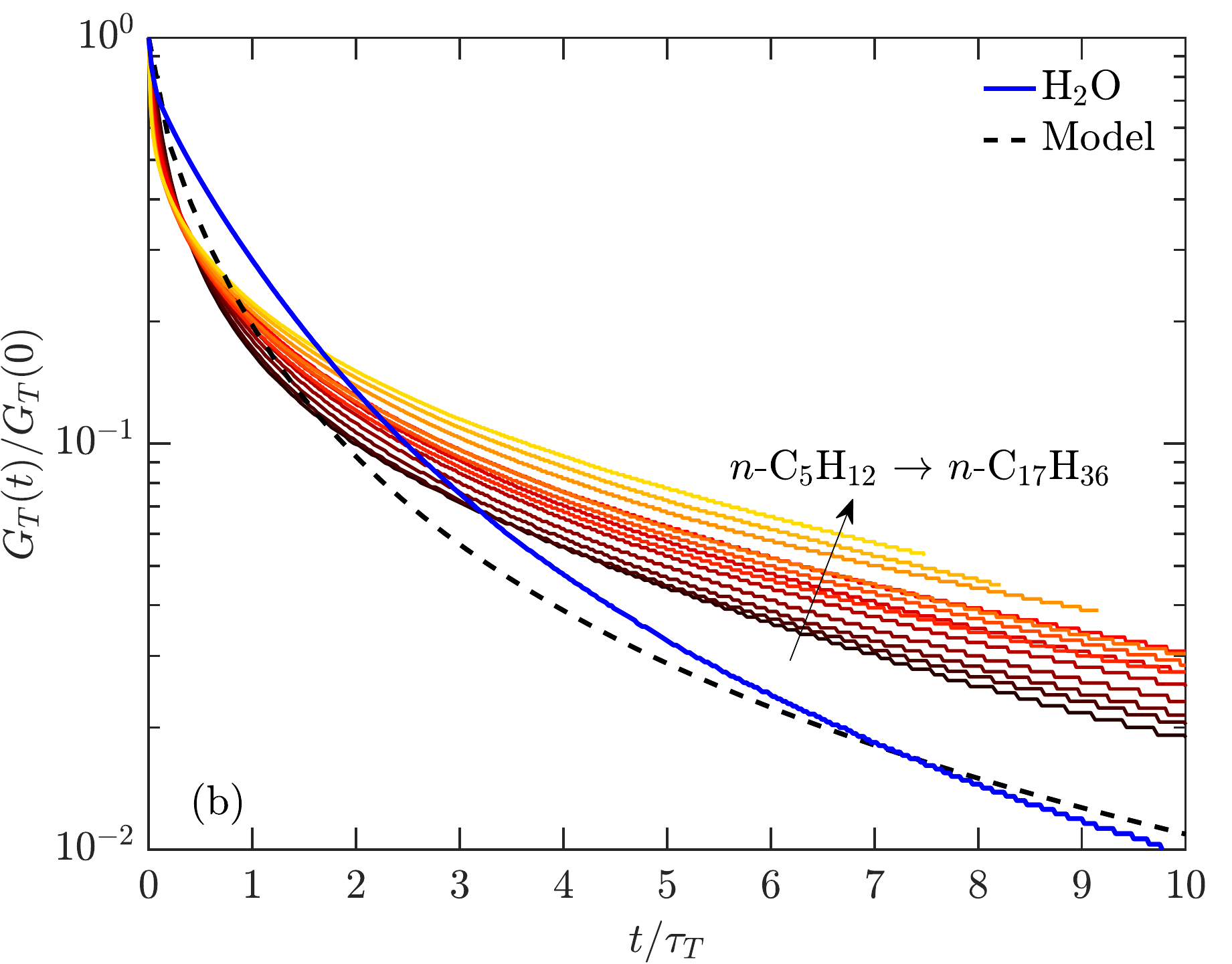} 
	\end{center}
	\caption{MD simulations of (a) autocorrelation function $G_R(t)$ for intramolecular interactions using Eq. \ref{eq:GmRT} and (b) autocorrelation function $G_T(t)$ for intermolecular interactions using Eq. \ref{eq:GmRT}, for bulk alkanes (solid lines, arrow shows increasing carbon number) and water (blue line), at $20^{\rm o}$C. $y$-axis has been normalized by zero time value $G_{R,T}(0)$, and $x$-axis has been normalized by correlation time $\tau_{R,T}$ (Eq. \ref{eq:TauRT}). The relaxation behavior based on the traditional hard-sphere model (Eq.~\ref{eq:Gmodel}) is indicated by the black dashed line.} 
	\label{fg:Gt}
\end{figure}
The simulation results for $G_{R,T}(t)$ are shown in Fig. \ref{fg:Gt}, where both $x$ and $y$ axes have been normalized for better comparison between the different fluids. The parameters used for the normalization are detailed below. A significant parameter in the analysis is the autocorrelation at $t = 0$, which is given by the following expression:
\begin{multline}
	G_{R,T}(0) = \frac{3}{16} \! \left(\frac{\mu_0}{4\pi}\right)^2 \! \hbar^2 \gamma^4  \frac{1}{N_{R,T}}\sum\limits_{i \neq j}^{N_{R,T}}  \\ 
	 \left< \frac{(3\cos^{2}\!\theta_{ij}\!(\tau)-1)^2}{r_{ij}^6\!\left(\tau\right)}  \right>_{\!\! \tau}
	\label{eq:GmRT0}
\end{multline}
NMR relaxation theory states a simple relation between $G_{R,T}(0)$ and the second-moment $\Delta\omega_{R,T}^2$ of the dipole-dipole interaction as such \cite{cowan:book}:
\begin{equation}
G_{R,T}(0) = \frac{1}{3}\Delta\omega_{R,T}^2,
\label{eq:G0SM}
\end{equation}
which leads to the final expression:
\begin{multline}
	\Delta\omega_{R,T}^2 = \frac{9}{16} \! \left(\frac{\mu_0}{4\pi}\right)^2 \! \hbar^2 \gamma^4  \frac{1}{N_{R,T}}\sum\limits_{i \neq j}^{N_{R,T}}  \\ 
	 \left< \frac{(3\cos^{2}\!\theta_{ij}\!(\tau)-1)^2}{r_{ij}^6\!\left(\tau\right)}  \right>_{\!\! \tau}
	\label{eq:SMRT}.
\end{multline}
The second-moment is significant because it determines the strength of the relaxation mechanism, and it is a strong function ($\Delta\omega_{R,T}^2 \propto r_{ij}^{-6}$) of the nuclear spin separation $r_{ij}$. When the fluctuations in $r_{ij}(t)$ and $\theta_{ij}(t)$ are uncorrelated, the angular term in the numerator can be averaged out, resulting in the following approximate expression:
\begin{equation}
	\Delta\omega_{R,T}^2 \approx \frac{9}{20} \! \left(\frac{\mu_0}{4\pi}\right)^2 \! \hbar^2 \gamma^4  \frac{1}{N_{R,T}}\sum\limits_{i \neq j}^{N_{R,T}}  \left< \frac{1}{r_{ij}^6\!\left(\tau\right)}  \right>_{\!\! \tau}
	\label{eq:SMRTuncorr}
\end{equation}
MD simulations indicate that Eq. \ref{eq:SMRTuncorr} holds within $\lesssim 0.1 \%$ for both intramolecular and intermolecular interactions, indicating that the ensemble averages of $r_{ij}(t)$ and $\theta_{ij}(t)$ are indeed uncorrelated. This approximation will not necessarily hold for fluids under confinement such as in nano-porous media, where complications from residual dipolar-coupling may come into play \cite{washburn:cmr2014}. In such cases the full expression in Eq. \ref{eq:SMRT} must be used.

Once the autocorrelation function $G_{R,T}(t)$ is determined from MD simulations, the spectral density of the local magnetic-field fluctuations is determined by Fourier transform:
\begin{equation}
J_{R,T}(\omega) = 2\int_{0}^{\infty}G_{R,T}(t)\cos\left(\omega t\right) dt,
\label{eq:FourierRTcos}
\end{equation}
given that $G_{R,T}(t)$ is real ($m = 0$) and an even function of $t$. $J_{R,T}(\omega)$ (in units of s$^{-1}$) is defined using the non-unitary angular-frequency form of the Fourier transform. The relaxation times $T_{1,R,T}$ and $T_{2,R,T}$ are then determined by $J_{R,T}(\omega)$ at specific angular frequencies given by \cite{mcconnell:book}
\begin{eqnarray}
\begin{aligned}
\frac{1}{T_{1,R,T}} &= J_{R,T}(\omega_0) + 4 J_{R,T}(2\omega_0),  \\
\frac{1}{T_{2,R,T}} &= \frac{3}{2} J_{R,T}(0) + \frac{5}{2} J_{R,T}(\omega_0) + J_{R,T}(2\omega_0) \, ,
\end{aligned}  
\label{eq:T12R}
\end{eqnarray}
where $\omega_0 = \gamma B_0$ is the Larmor frequency, and $\gamma/2\pi = 42.58$ MHz/T is the $^1$H nuclear gyro-magnetic ratio. 

The relaxation rates are additive, giving the following final expression:
\begin{eqnarray}
\begin{aligned}
\frac{1}{T_{1}} &= \frac{1}{T_{1,R}} + \frac{1}{T_{1,T}}, \\
\frac{1}{T_{2}} &= \frac{1}{T_{2,R}} + \frac{1}{T_{2,T}} \\
\label{eq:T12sum}
\end{aligned} 
\end{eqnarray}
which are the measurable quantities in the laboratory.

The most significant parameter in the description of NMR relaxation in liquids is the correlation time $\tau_{R,T}$, which is defined by the following expressions \cite{cowan:book}:
\begin{equation}
\tau_{R,T} = \frac{1}{G_{R,T}(0)}\int_{0}^{\infty}G_{R,T}(t) dt = \frac{1}{2}\frac{J_{R,T}(0)}{G_{R,T}(0)}.
\label{eq:TauRT}
\end{equation}
In other words, $\tau_{R,T}$ is defined as the normalized area under the autocorrelation functions $G_{R,T}(t)$. Eq. \ref{eq:TauRT} also shows the simple relation to the spectral density $J_{R,T}(0)$ at $\omega = 0$ using Eq. \ref{eq:FourierRTcos}. 


As shown below in Section \ref{sc:Results}, the correlation times $\tau_{R,T}$ for the bulk fluids are in the fast motion regime, i.e. $\omega_0 \tau_{R} \ll 1$ and $\omega_0 \tau_{T} \ll 1$, where $\omega_0/2\pi = 2 $ MHz for the NMR relaxation measurements \cite{shikhov:amr2016} used in this report. In the fast motion regime, the spectral density $J_{R,T}(\omega)$ becomes independent of $\omega$ for $\omega \lesssim 2 \omega_0$, and equal to the $\omega = 0$ component as such:
\begin{equation}
J_{R,T}(0) = J_{R,T}(\omega_0) = J_{R,T}(2\omega_0) = \frac{2}{3} \Delta\omega_{R,T}^2 \tau_{R,T}. 
\label{eq:JRTmotional}
\end{equation}
From Eqs.~\ref{eq:T12R} and~\ref{eq:JRTmotional} we obtain the following expression for the fast motion regime:
\begin{equation}
\frac{1}{T_{1,R,T}} = \frac{1}{T_{2,R,T}} = 5J_{R,T}(0) = \frac{10}{3} \Delta\omega_{R,T}^2  \tau_{R,T}.
\label{eq:T12RTmotional}
\end{equation}
In other words, $T_{1,R} = T_{2,R}$ and $T_{1,T} = T_{2,T}$, which is characteristic for low viscosity fluids. 
From Eqs.~Eq. \ref{eq:T12RTmotional} and~\ref{eq:T12sum} we thus obtain
\begin{equation}
\frac{1}{T_1} = \frac{1}{T_2} =\frac{10}{3} \Delta\omega_R^2  \tau_R +  \frac{10}{3} \Delta\omega_T^2 \tau_T,
\label{eq:T12motional}
\end{equation}
which is the final expression used to predict the NMR relaxation times from simulation results. 

As shown below, the MD simulation results are compared with measured (de-oxygenated) $T_{2}$ data from Shikhov {\it et al.} \cite{shikhov:amr2016}. The NMR measurements were acquired at $\omega_0/2\pi = 2$ MHz using a CPMG echo train with an echo spacing of $T_E = $ 0.2 ms. The samples were carefully de-oxygenated by the freeze-pump-thaw technique in order to remove unwanted (in the present case) NMR relaxation from paramagnetic O$_2$ molecules in solution. 


\subsection{Diffusion}\label{sc:Diffusion}

\begin{figure}[h!]
	\begin{center}
		\includegraphics[width=0.9\columnwidth]{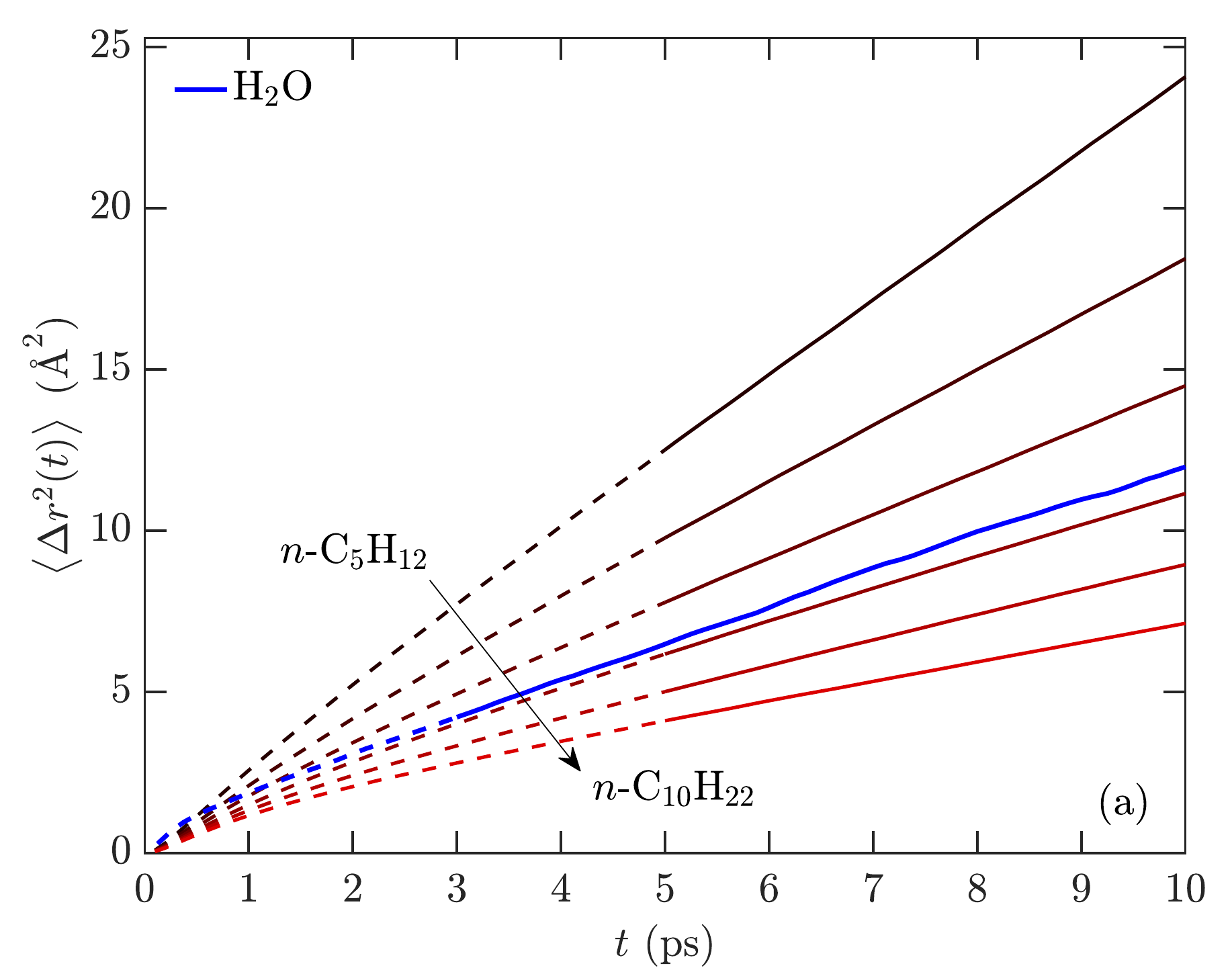} 
		\includegraphics[width=0.9\columnwidth]{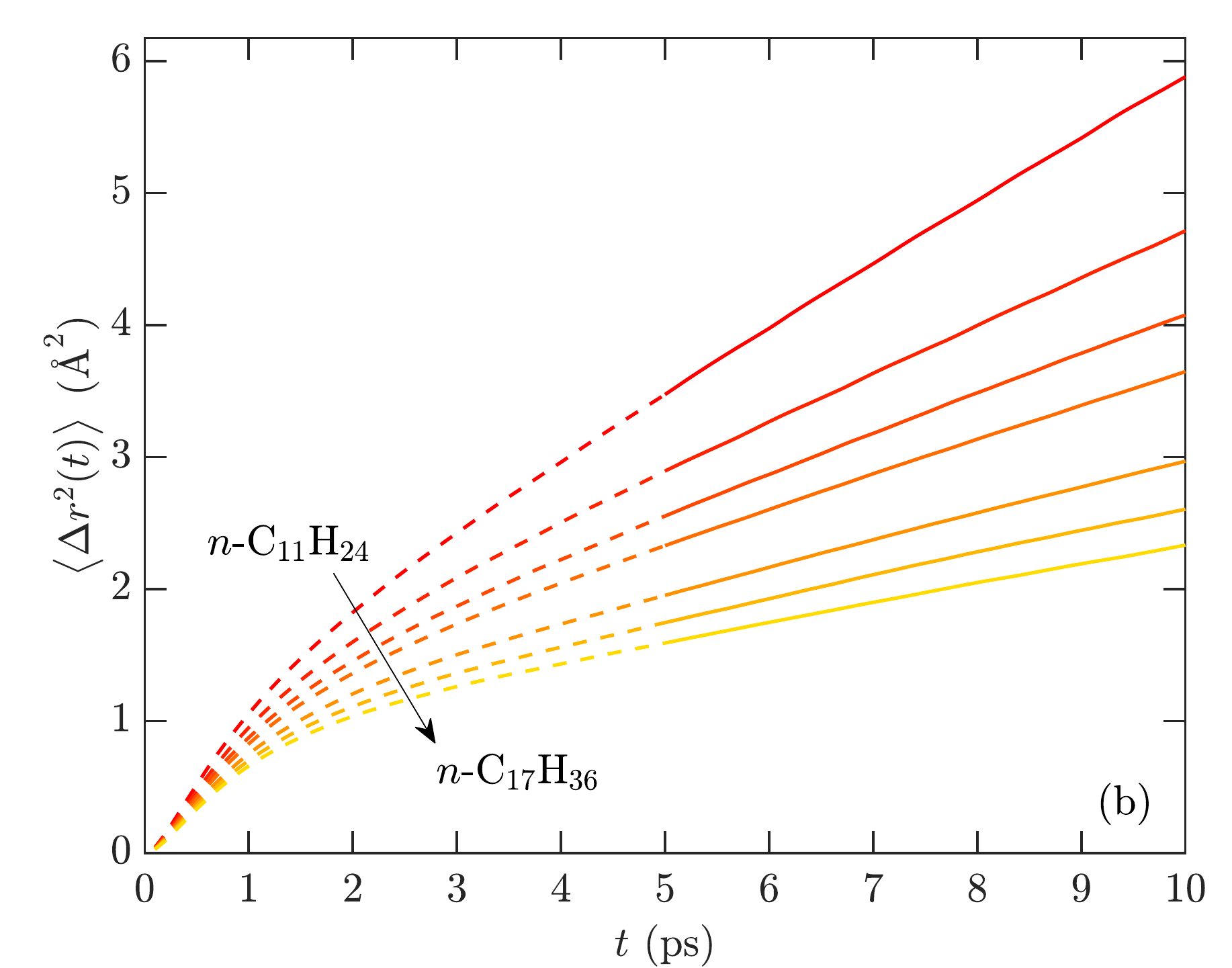} 
	\end{center}
	\caption{MD simulations of mean-square displacement $\Delta r^2 (t)$ versus time $t$ for (a) water and $n$-C$_{5}$H$_{12}$ $\rightarrow$ $n$-C$_{10}$H$_{22}$ (arrow shows increasing carbon number), and (b) $n$-C$_{11}$H$_{24}$ $\rightarrow$ $n$-C$_{17}$H$_{36}$, at $20^{\rm o}$C. Solid lines show fitting region used to obtain translational diffusion coefficient $D_T$ from Eq. \ref{eq:Dsimulation} for $t > 5 $ ps, and dashed lines show early time regime not used in the fit.} 
	\label{fg:Diff}
\end{figure}

An independent computation of translational diffusion $D_T$ was performed from MD simulations. The mean square displacement was computed from the center-of-mass of the molecule and plotted against diffusion evolution time $t ( < 10)$ ps (Fig.~\ref{fg:Diff}). 

At long-times ($t$), the slope of the linear diffusive regime (Fig.~\ref{fg:Diff}) gives the translational self-diffusion coefficient $D_T$ according to Einstein's relation:
\begin{equation}
\frac{d \left<\Delta r^2\right>}{d t} = 6 D_T 
\label{eq:Dsimulation}
\end{equation}
In the linear regression procedure, the early ballistic regime and part of the linear regime is excluded to obtain a robust estimate of $D_T$.

As shown below, the MD simulation results are compared with measured $D_T$ data acquired using $^1$H NMR diffusivity, where diffusion was encoded using uni-polar pulsed-field-gradient stimulated-echo experiments \cite{mitchell:pnmrs2014} with diffusion evolution times of the order $\Delta \gtrsim 10$ ms. Due to an incomplete range of alkanes $n$-C$_5$H$_{12}$ $\rightarrow$ $n$-C$_{17}$H$_{36}$ from a single measurement source, the $D_T$ measurements were taken from Shikhov {\it et al.} \cite{shikhov:amr2016} at $\omega_0/2\pi = 2 $ MHz, and from Tofts {\it et al.} \cite{tofts:mrm2000} at $\omega_0/2\pi = 200 $ MHz.

\section{Results and Discussions}\label{sc:Results}

This section presents the results of the NMR relaxation and diffusion, followed by analysis using hard-sphere models. 

\subsection{NMR Relaxation}

\begin{figure}[h!]
	\begin{center}
		\includegraphics[width=0.9\columnwidth]{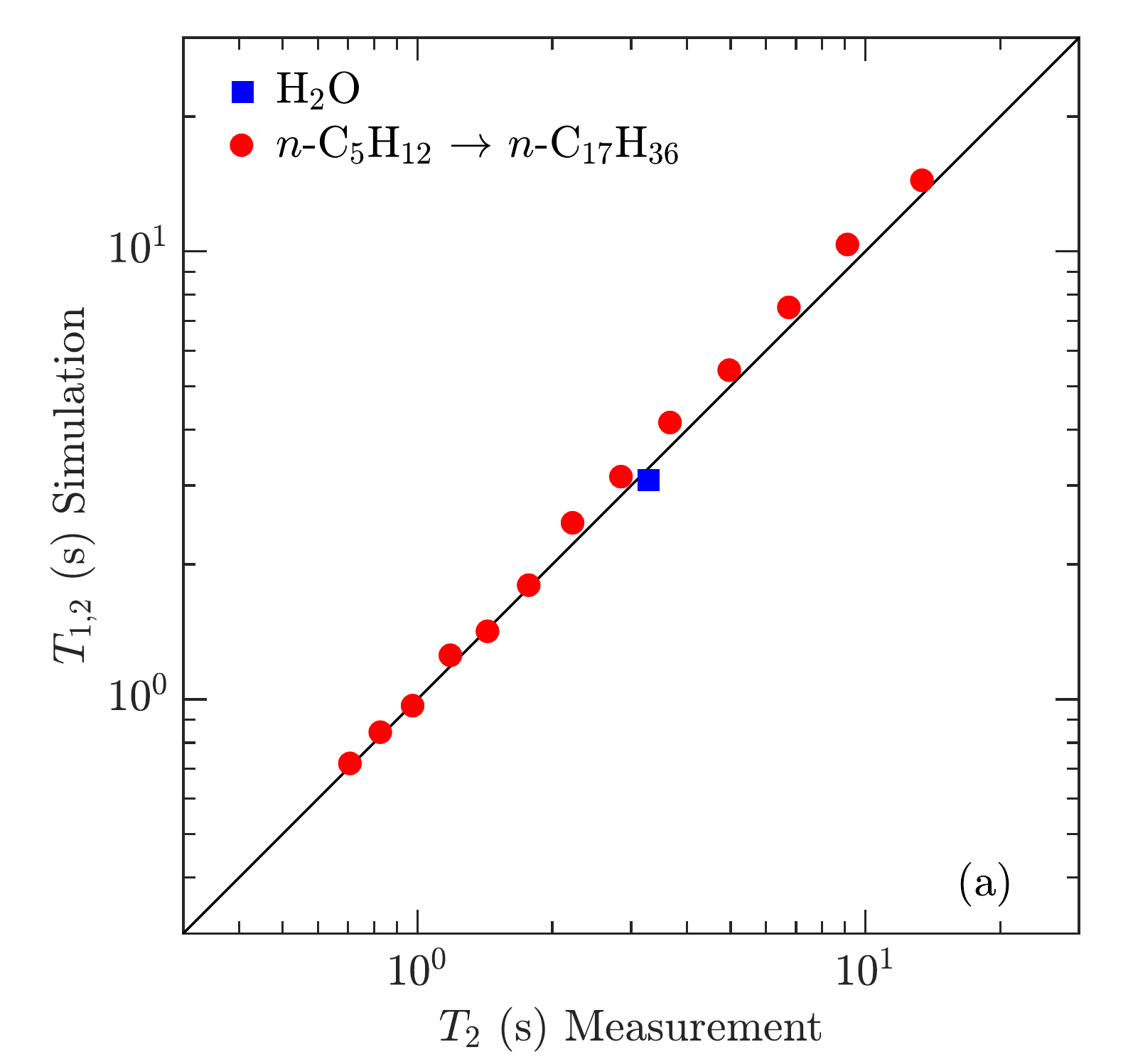}
		\includegraphics[width=0.9\columnwidth]{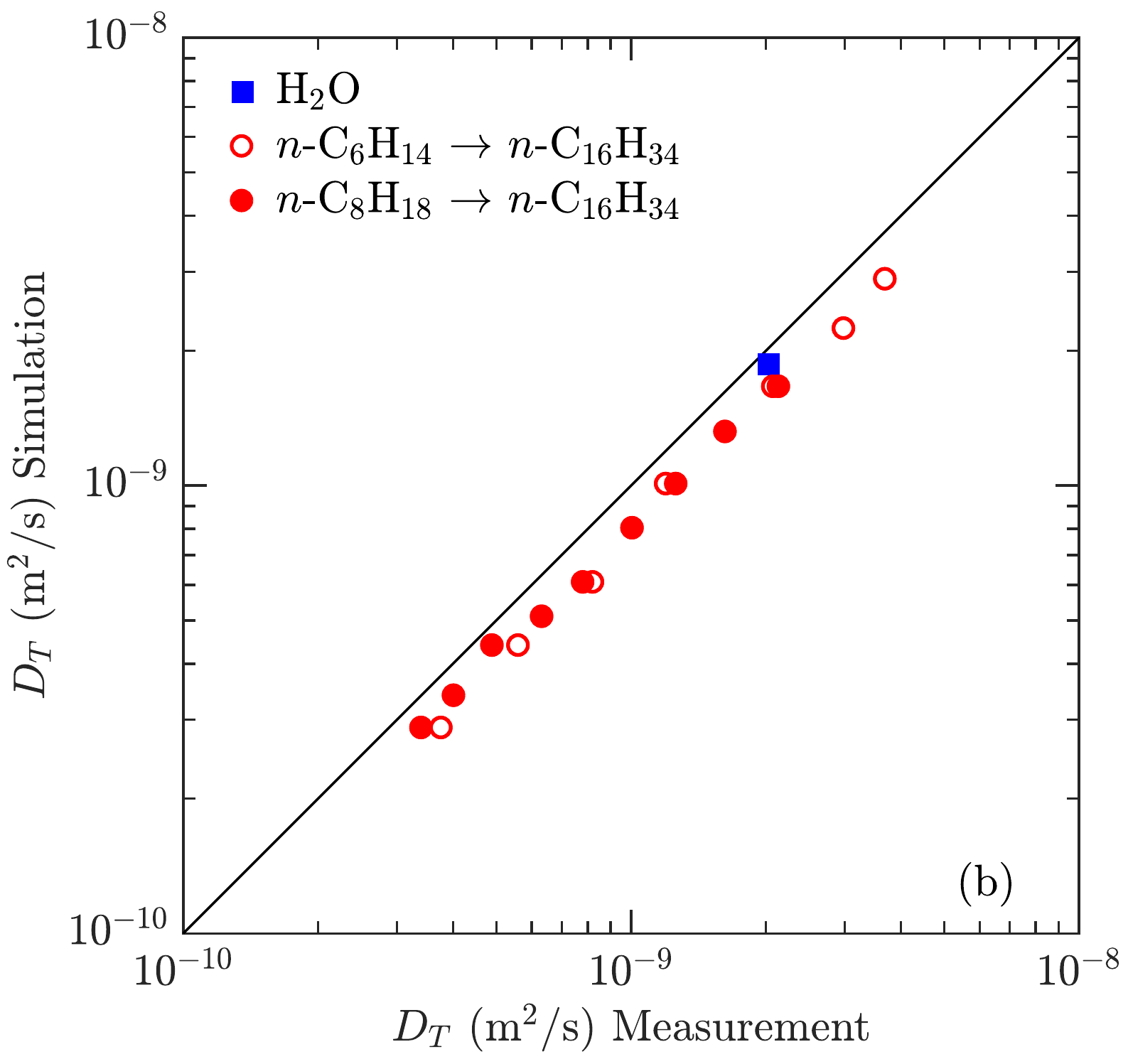} 			
	\end{center}
	\caption{(a) Cross-plot of predicted NMR relaxation time $T_{1,2}$ (Eq. \ref{eq:T12motional}) from MD simulations on the $y$-axis versus (de-oxygenated) $T_2$ measurements taken from Shikhov {\it et al.} \cite{shikhov:amr2016} on the $x$-axis, for alkanes and water, showing a very strong correlation of $R^2 = 0.998$ and a small deviation of $nrmsd = 7.5\%$ (Eq. \ref{eq:nrmsd}). (b) Cross-plot of predicted translational diffusion coefficient $D_T$ (Eq. \ref{eq:Dsimulation}) from MD simulations on the $y$-axis versus $D_T$ measurements taken from Shikhov {\it et al.} \cite{shikhov:amr2016} (open symbols) and taken from Tofts {\it et al.} \cite{tofts:mrm2000} (closed symbols) on the $x$-axis, for alkanes and water, showing a strong correlation of $R^2 = 0.988$ and a deviation of $nrmsd = 19.8\%$. Data are at $20^{\rm o}$C (or $22.6^{\rm o}$C), and are listed in Table \ref{tb:1}. Carbon number increases from right-to-left in (a) and (b).}
	\label{fg:Xplot}
\end{figure}
\begin{table*}
	\begin{center}
		\begin{tabularx}{18cm}{CCCCCCCCCCCCC}
			\hline
			
		\multicolumn{2}{c}{$^{{\strut}{}}$Fluid}  &$T_2$  &  $T_{1,2}$ & $T_{1,2,R}$ & $T_{1,2,T}$ & $\tau_R$ & $\tau_T$ & $\Delta\omega_R$& $\Delta\omega_T$& $D_T$& $D_T$ & $D_T$   \\ 
			$^{{\strut}{}}$& & de-oxy &  &  &  &    & &/2$\pi$ & /2$\pi$& ($10^{-9}$& ($10^{-9}$ & ($10^{-9}$ \\ 
			$^{{\strut}{}}$& & (s) & (s) & (s) & (s) & (ps) & (ps) &  (kHz)& (kHz) & m$^2$/s)& m$^2$/s) & m$^2$/s) \\ 
			$^{{\strut}{}}$& & Ref. \cite{shikhov:amr2016} & MD & MD  & MD & MD & MD & MD& MD & Ref. \cite{shikhov:amr2016} & Ref. \cite{tofts:mrm2000} & MD \\
			$^{{\strut}{}}$& & 22.6$^{\rm o}$C& 20$^{\rm o}$C& 20$^{\rm o}$C & 20$^{\rm o}$C & 20$^{\rm o}$C & 20$^{\rm o}$C & 20$^{\rm o}$C & 20$^{\rm o}$C & 22.6$^{\rm o}$C& 20$^{\rm o}$C & 20$^{\rm o}$C\\ 
			\hline
			
\multicolumn{2}{c}{$^{{\strut}{}}$H$_2$O}	&	3.28	&	3.09	&	5.22	&	7.56	&	2.70	&	3.99	&	23.22	&	15.87	&	2.15	&	2.03	&	1.87	\\
\multicolumn{2}{c}{$n^{{\strut}{}}$-C$_{5}$H$_{12}$}	&	13.44	&	14.34	&	24.26	&	35.06	&	0.77	&	2.32	&	20.17	&	9.66	&		&		&	3.85	\\
\multicolumn{2}{c}{$n^{{\strut}{}}$-C$_{6}$H$_{14}$}	&	9.16	&	10.31	&	16.38	&	27.81	&	1.15	&	2.89	&	20.05	&	9.72	&	3.70	&		&	2.88	\\
\multicolumn{2}{c}{$n^{{\strut}{}}$-C$_{7}$H$_{16}$}	&	6.78	&	7.46	&	11.13	&	22.63	&	1.71	&	3.55	&	19.99	&	9.73	&	2.99	&		&	2.24	\\
\multicolumn{2}{c}{$n^{{\strut}{}}$-C$_{8}$H$_{18}$}	&	4.99	&	5.40	&	7.64	&	18.47	&	2.50	&	4.35	&	19.93	&	9.73	&	2.08	&	2.14	&	1.66	\\
\multicolumn{2}{c}{$n^{{\strut}{}}$-C$_{9}$H$_{20}$}	&	3.68	&	4.13	&	5.67	&	15.22	&	3.40	&	5.24	&	19.87	&	9.76	&		&	1.63	&	1.31	\\
\multicolumn{2}{c}{$n^{{\strut}{}}$-C$_{10}$H$_{22}$}	&	2.86	&	3.13	&	4.14	&	12.72	&	4.66	&	6.30	&	19.83	&	9.74	&	1.20	&	1.26	&	1.01	\\
\multicolumn{2}{c}{$n^{{\strut}{}}$-C$_{11}$H$_{24}$}	&	2.23	&	2.47	&	3.20	&	10.76	&	6.05	&	7.45	&	19.82	&	9.74	&		&	1.01	&	0.80	\\
\multicolumn{2}{c}{$n^{{\strut}{}}$-C$_{12}$H$_{26}$}	&	1.78	&	1.79	&	2.35	&	7.55	&	8.27	&	10.62	&	19.80	&	9.73	&	0.82	&	0.78	&	0.61	\\
\multicolumn{2}{c}{$n^{{\strut}{}}$-C$_{13}$H$_{28}$}	&	1.44	&	1.41	&	1.80	&	6.50	&	10.79	&	12.42	&	19.76	&	9.70	&		&	0.63	&	0.51	\\
\multicolumn{2}{c}{$n^{{\strut}{}}$-C$_{14}$H$_{30}$}	&	1.19	&	1.25	&	1.58	&	5.95	&	12.40	&	13.66	&	19.71	&	9.66	&	0.56	&	0.49	&	0.44	\\
\multicolumn{2}{c}{$n^{{\strut}{}}$-C$_{15}$H$_{32}$}	&	0.98	&	0.96	&	1.20	&	4.94	&	16.29	&	16.37	&	19.75	&	9.69	&		&	0.40	&	0.34	\\
\multicolumn{2}{c}{$n^{{\strut}{}}$-C$_{16}$H$_{34}$}	&	0.83	&	0.84	&	1.04	&	4.44	&	18.96	&	18.24	&	19.66	&	9.69	&	0.38	&	0.34	&	0.29	\\
\multicolumn{2}{c}{$n^{{\strut}{}}$-C$_{17}$H$_{36}$}	&	0.71	&	0.72	&	0.87	&	4.12	&	22.41	&	20.05	&	19.79	&	9.59	&		&		&	0.25	\\

			\hline
		\end{tabularx}
	\end{center}
	\caption{Simulation and measurements of NMR relaxation and diffusion, including: de-oxygenated NMR relaxation time $T_2$ from measurements at $\omega_0/2\pi = 2 $ MHz taken from Shikhov {\it et al.} \cite{shikhov:amr2016}, total $T_{1,2}$ from MD simulations (Eq. \ref{eq:T12motional}), intramolecular component $T_{1,2,R}$ from MD simulations (Eq. \ref{eq:T12RTmotional}), intermolecular component $T_{1,2,T}$ from MD simulations (Eq. \ref{eq:T12RTmotional}), intramolecular correlation time $\tau_R$ from MD simulations (Eq. \ref{eq:TauRT}), intermolecular correlation time $\tau_T$ from MD simulations (Eq. \ref{eq:TauRT}), square-root of intramolecular second-moment $\Delta\omega_R/2\pi$ from MD simulations (Eq. \ref{eq:SMRT}), square-root of intermolecular second-moment $\Delta\omega_T/2\pi$ from MD simulations (Eq. \ref{eq:SMRT}), translational diffusion coefficient $D_T$ from NMR measurements at $\omega_0/2\pi = 2 $ MHz taken from Shikhov {\it et al.} \cite{shikhov:amr2016}, $D_T$ from NMR measurements at $\omega_0/2\pi = 200 $ MHz taken from Tofts {\it et al.} \cite{tofts:mrm2000}, $D_T$ from MD simulations (Eq. \ref{eq:Dsimulation}). The temperature of each entry is also listed.
		\label{tb:1} }
\end{table*}

The results and analysis leading up to Eq. \ref{eq:T12motional} do not have any adjustable parameters in the interpretation of the simulation data. As shown in Fig. \ref{fg:Xplot}(a), the MD simulation results are very strongly correlated with measurement ($R^2 = 0.998$), and an estimate of the deviation can be established using the normalized root-mean-square deviation $nrmsd$, defined as
\begin{equation}
nrmsd = \sqrt{\frac{1}{N}\sum\limits_{i=1}^{N}\left(\frac{Y_i - X_i}{X_i}\right)^2 }\times 100,
\label{eq:nrmsd}
\end{equation}
where $Y_i$ are the MD simulation results, $X_i$ are the measurements, and $N$ are the number of data points. The data in Fig. \ref{fg:Xplot}(a) indicate a small $nrmsd \sim 7.5 \%$, which could come from uncertainties in the simulations and/or the measurements. The slight difference in temperature between simulation (20$^{\rm o}$C) and measurement (22.6$^{\rm o}$C) is unlikely to be the cause of the deviation, since this would result in a systematic difference for all the data. The experimental challenge is also to fully de-oxygenate the samples; imperfect de-oxygenation would cause the measured $T_2$ to be systematically lower than $T_{1,2}$ from simulation, however this is also not found to be the case. On the computational side, the most likely source of uncertainties is from the truncation of $G_{R,T}(t)$ before $G_{R,T}(t)\rightarrow 0$. This has the effect of underestimating $\tau_{R,T}$ (Eq. \ref{eq:TauRT}) and therefore overestimating $T_{1,2}$ (Eq. \ref{eq:T12motional}). 

\begin{figure}
	\begin{center}
		\includegraphics[width=0.9\columnwidth]{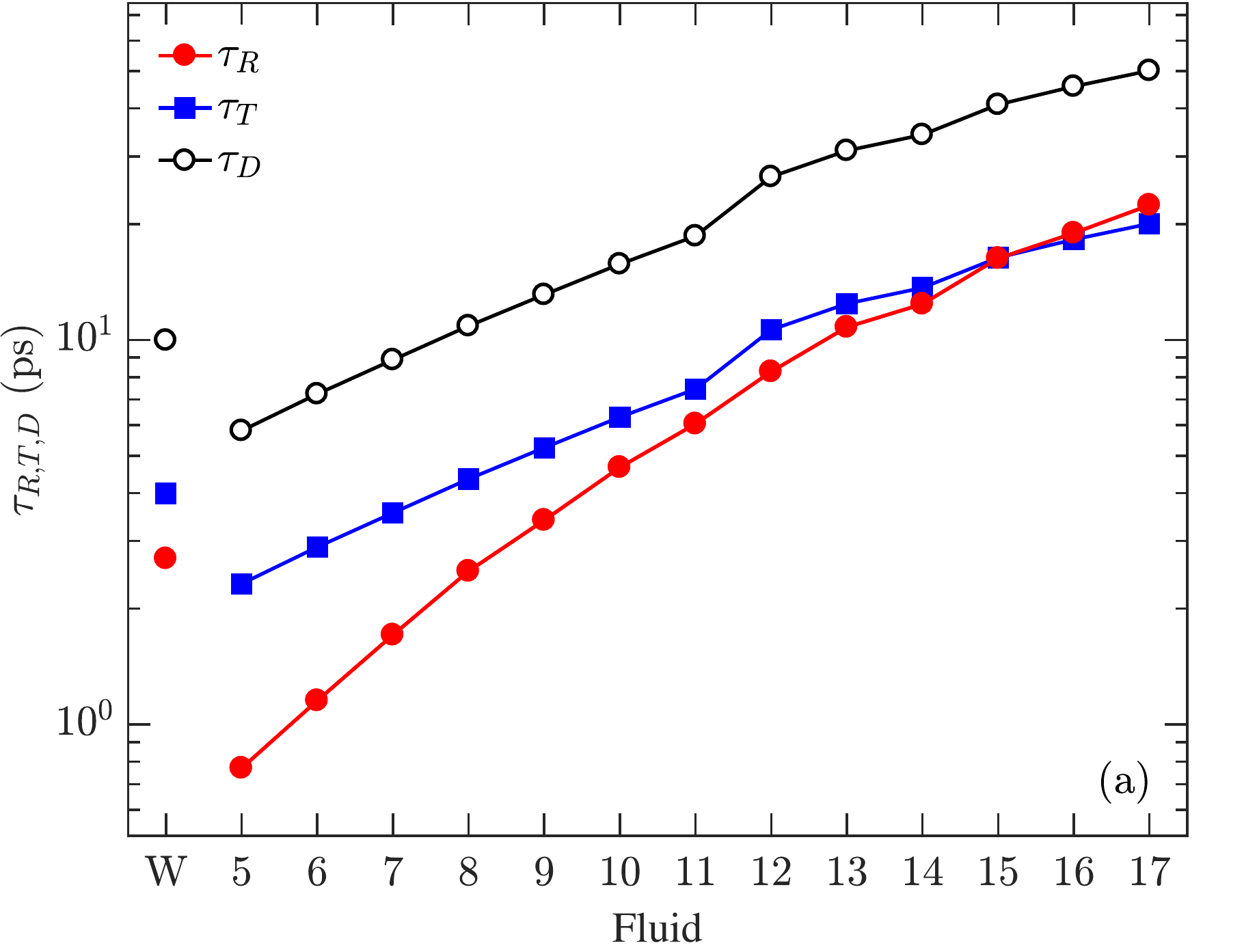} 
		\includegraphics[width=0.9\columnwidth]{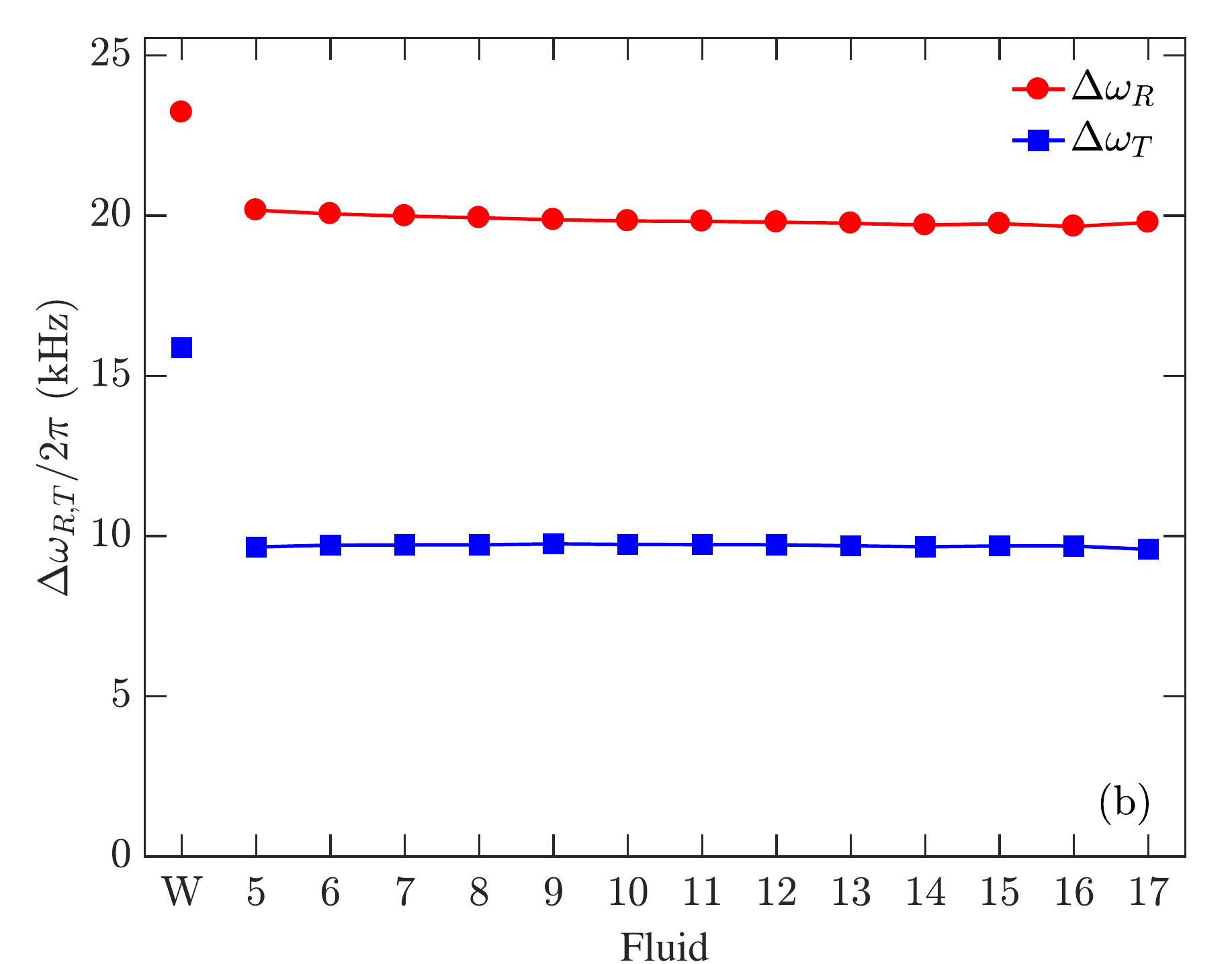} 
	\end{center}
	\caption{(a) MD simulation result for intramolecular correlation time $\tau_R$ (Eq. \ref{eq:TauRT}), intermolecular correlation time $\tau_T$ (Eq. \ref{eq:TauRT}), and intermolecular diffusion correlation time $\tau_D (= \frac{5}{2} \tau_T)$, for water (W) and alkanes (carbon number C5$\rightarrow$C17). (b) MD simulation result for square-root of intramolecular second moment $\Delta\omega_R$ (Eq. \ref{eq:SMRT}) and square-root of intermolecular second moment $\Delta\omega_T$ (Eq. \ref{eq:SMRT}), for water (W) and alkanes (C5$\rightarrow$C17). Data are at $20^{\rm o}$C, and are listed in Table \ref{tb:1}.} \label{fg:TauSM}
\end{figure}

The resulting correlation times $\tau_{R,T}$ are listed in Table \ref{tb:1}, and plotted in Fig. \ref{fg:TauSM}(a). For the alkanes, both $\tau_{R,T}$ increase by over an order of magnitude with increasing chain length (i.e. carbon number). $\tau_{R}$ is also found to be $\approx 3\times$ shorter than $\tau_{T}$ at low carbon number, and are found to merge at high carbon number. The $\tau_{R,T}$ values for water are found to be comparable to octane ($n$-C$_{8}$H$_{18}$). 
The resulting second-moments $\Delta\omega_{R,T}^2$ are listed in Table \ref{tb:1} and plotted in Fig. \ref{fg:TauSM}(b). The data are presented as the square-root $\Delta\omega_{R,T}/2\pi$ ($\propto r_{ij}^{-3}$) in units of kHz. For the alkanes, the intramolecular $\Delta\omega_{R}/2\pi \simeq 20$ kHz are found to be $\approx 2\times$ larger than the intermolecular $\Delta\omega_{T}/2\pi \simeq 10$ kHz. The values for intermolecular
contribution to the second moment for water are $\approx 50 \%$ larger than alkanes, which is expected given the closer packing of water due to hydrogen bonding.


As shown in Fig. \ref{fg:Xplot}(a), the MD simulations provide an accurate method of predicting the total NMR relaxation times $T_{1,2}$. The next natural step is to use the MD simulations to obtain new insight into intramolecular ${T_{1,2,R}}$ versus intermolecular ${T_{1,2,T}}$ relaxation. The first important insight is the relative strength of the two mechanisms. Fig. \ref{fg:Ratios}(a) shows the ratio ${T_{1,2,T}}/{T_{1,2,R}}$ for each fluid, where a large ratio indicates that the intramolecular relaxation (short ${T_{1,2,R}}$) dominates over intermolecular (long ${T_{1,2,T}}$). Both pentane and water indicate that ${T_{1,2,T}}/{T_{1,2,R}} \simeq1.5$, however the ratio steadily increases with increasing chain length. The data suggests that larger molecules become increasingly dominated by intramolecular interactions (i.e. rotational motions) versus intermolecular interactions (i.e. translational motions). This is an important finding for simulating NMR relaxation in macromolecules such as polymers and proteins, where,
perhaps without much justification,  the common practice is to only consider intramolecular dipole-dipole interactions.

\subsection{Diffusion}

The MD simulation results for $D_T$ are listed in Table \ref{tb:1} and plotted in Fig. \ref{fg:Xplot}(b) against measurements. The NMR diffusion measurement is (in principle) independent of the applied magnetic field $\omega_0$, provided no internal gradients exist, which is the case for bulk liquids. The NMR diffusion measurement is also (in principle) independent of whether the samples are air-saturated or de-oxygented. The air-saturated $D_T$ data were chosen from \cite{shikhov:amr2016} due to the larger range of alkanes ($n$-C$_{6}$H$_{14}$ $\rightarrow$ $n$-C$_{16}$H$_{34}$) measured. One potential complication in measuring NMR diffusion for bulk liquids is convection due to temperature gradients in the sample, which would tend to overestimate $D_T$. However, these measurements were taken at a stable ambient temperature and therefore should not contain such complications. 

While the MD simulations and measurements are strongly correlated ($R^2 = 0.988$), there is a systematic deviation of $nrmsd = 19.8 \%$ (Eq. \ref{eq:nrmsd}). Both the MD simulations and the measurements have statistical uncertainties associated with them, however it is evident from Fig. \ref{fg:Xplot}(b) that the deviation $nrmsd = 19.8 \%$ is dominated by a systematic offset, where MD simulations appear lower than measurement. It should be noted however that the agreement is remarkable given the huge difference in diffusion evolution time between MD simulation $t \simeq 10 $ ps and NMR measurement $\Delta \simeq 10$ ms. Put otherwise, the translational diffusion coefficient is compared over 9 orders of magnitude in diffusion evolution time, and found to be constant within $\sim 20 \%$. This could be viewed as a testament to Einstein's theory of Brownian motion (Eq. \ref{eq:Dsimulation}).


\subsection{Hard-Sphere Models}\label{cs:Hard}
In all of the results summarized in Table \ref{tb:2}, there are no models or adjustable parameters used in going from MD simulation results to predictions of NMR relaxation and diffusion. However in order to gain further insight into the molecular dynamics of bulk hydrocarbons and water, a physical model is required. 

The intramolecular dipole-dipole interaction \cite{bloembergen:pr1948} is traditionally derived using the rotational diffusion equation of rank 2 for hard-spheres of radius $R_R$, and is parameterized by the rotational diffusion coefficient $D_R$. The parameter $D_R$ is simply related to rotational correlation time $\tau_R$, which is defined as the average time it takes the molecule to rotate by 1 radian. The intermolecular dipole-dipole interaction \cite{torrey:pr1953} is traditionally derived using the Brownian motion model where the diffusion propagator is derived for hard-spheres of radius $R_T$, and is parameterized by the translational diffusion coefficient $D_T$. The parameter $D_T$ is simply related to the diffusion correlation time $\tau_D$, which is defined as the average time it takes the molecule to diffuse by one hard-core diameter $2R_T$ \cite{mcconnell:book}. 
For hard-spheres, the diffusion coefficients (and correlation times) can be related to the bulk properties of the fluid, namely the viscosity $\eta$ and absolute temperature $T$, using the traditional Stokes-Einstein relation \cite{abragam:book}:
\begin{eqnarray}
\begin{aligned}
\tau_R &= \frac{1}{6D_R} = \frac{4\pi }{3k_B} R_R^3 \frac{\eta}{T}, \\
\tau_D &=  \frac{2R_T^2}{D_T} = \frac{12\pi}{k_B} R_T^3 \frac{\eta}{T}= \frac{5}{2}\tau_T.
\end{aligned}  \label{eq:TauModel}
\end{eqnarray}
Note that in order to relate the Stokes-Einstein diffusion correlation time $\tau_D$ to the NMR derived translational correlation time $\tau_T$ (Eq. \ref{eq:TauRT}), a factor of $\frac{5}{2}$ is required, i.e. $\tau_D = \frac{5}{2}\tau_T$ \cite{cowan:book}. An expression for the two autocorrelation functions are then derived as such:
\begin{eqnarray}
\begin{aligned}
G_R(t) &= G_R(0) \exp\left(-\frac{t}{\tau_R}\right), \\
G_T(t) &= G_T(0) \int_{0}^{\infty}3 \frac{J_{3/2}^2(x)}{x}\exp\left(-x^2\frac{t}{\frac{5}{2}\tau_T}\right) dx
\end{aligned}  \label{eq:Gmodel}
\end{eqnarray}
where both expressions reduce to $G_{R,T}(0)$ (Eq. \ref{eq:G0SM}) at $t = 0$. 

\begin{figure}
	\begin{center}
		\includegraphics[width=0.9\columnwidth]{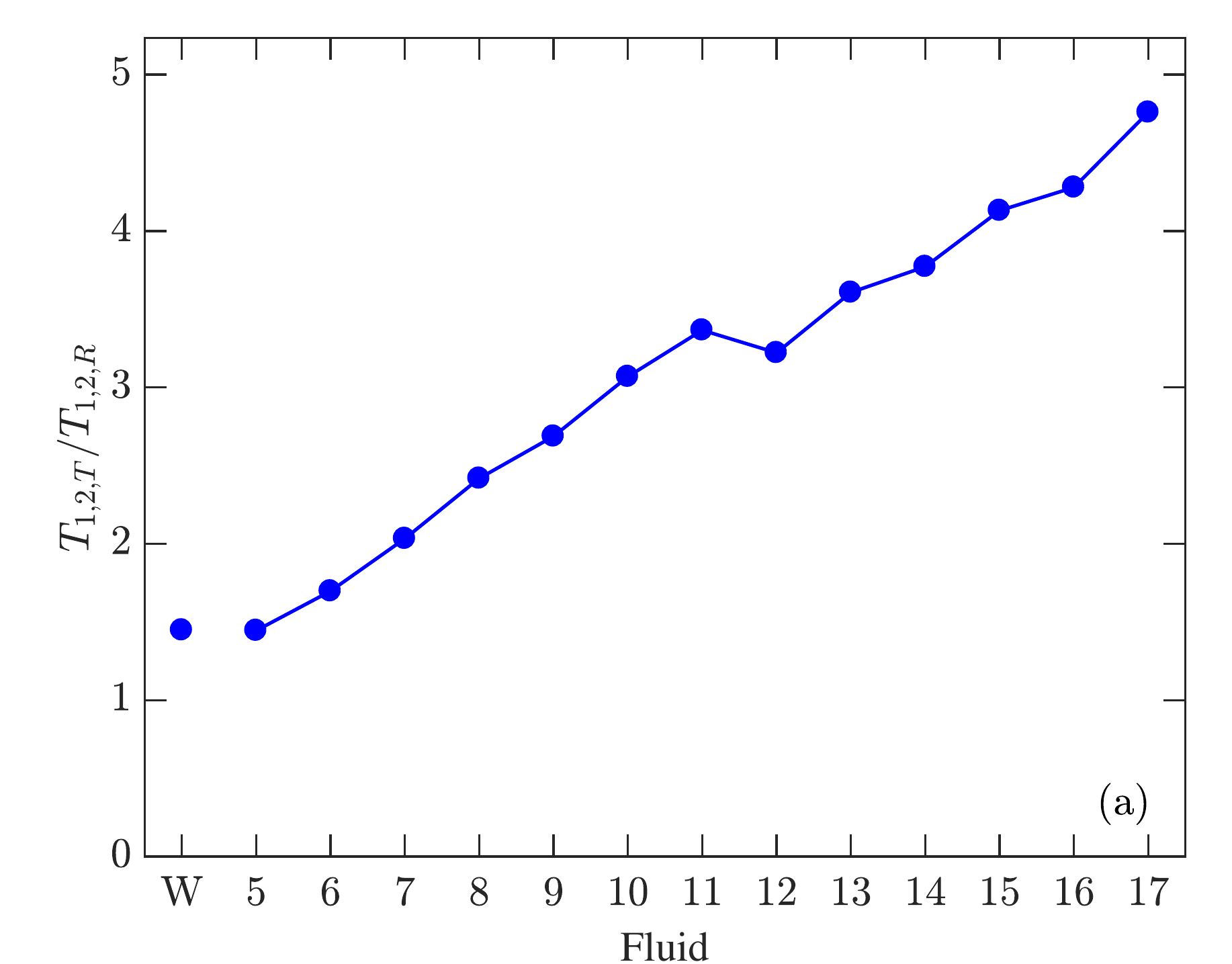} 
		\includegraphics[width=0.9\columnwidth]{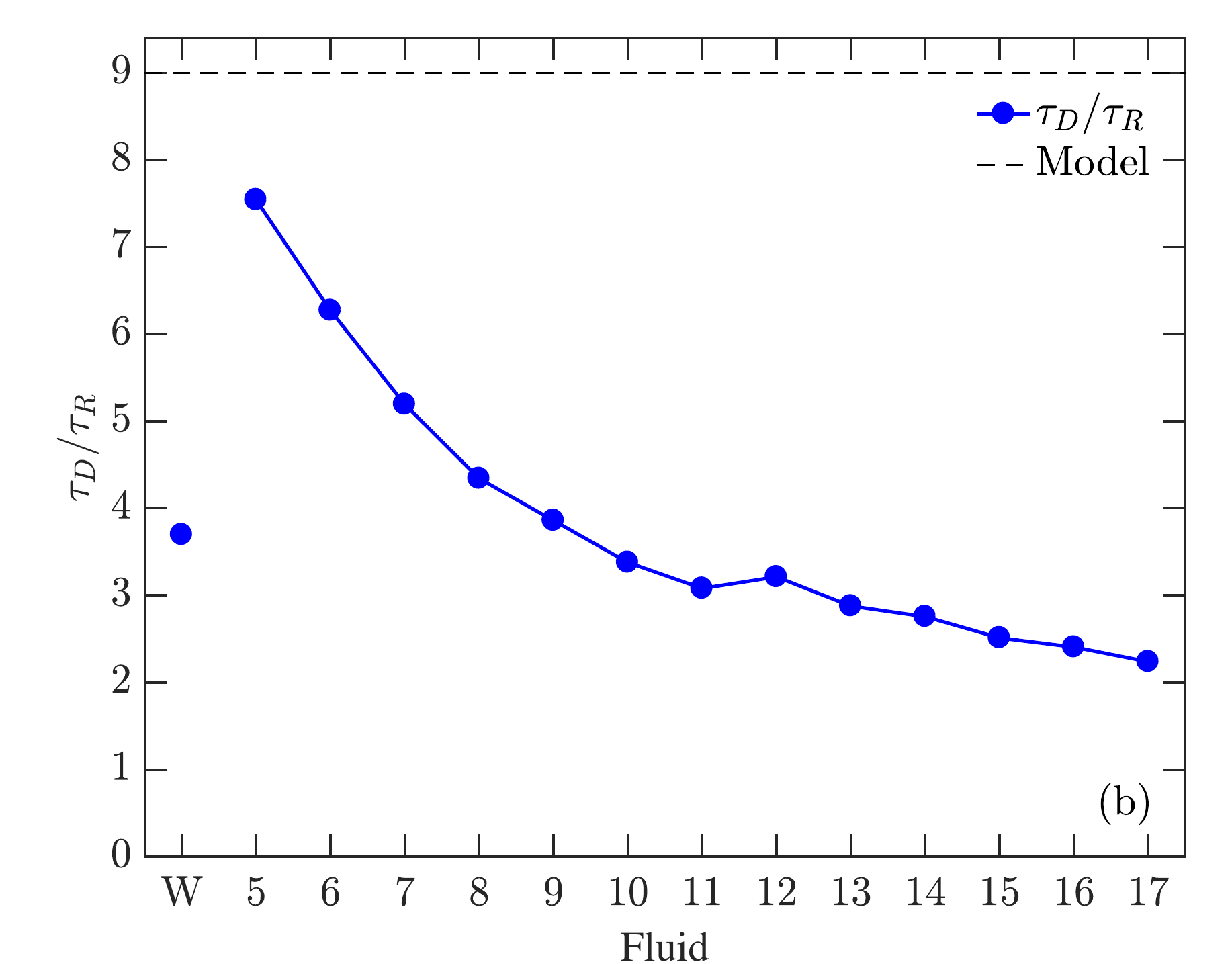}
	\end{center}
	\caption{(a) MD simulation result for ratio $T_{1,2,T}/T_{1,2,R}$ (Eq. \ref{eq:T12RTmotional}) of intermolecular relaxation time to intramolecular relaxation time, for water (W) and alkanes (carbon number C5$\rightarrow$C17). (b) MD simulation result for ratio $\tau_D/\tau_R$ of intermolecular diffusion correlation time ($\tau_D = \frac{5}{2}\tau_T$) to intramolecular correlation time ($\tau_R$), along with predicted value $\tau_D/\tau_R = 9$ \cite{abragam:book} from Stokes-Einstein relation (Eq. \ref{eq:TauModel}), for water (W) and alkanes (C5$\rightarrow$C17). Data are at $20^{\rm o}$C, and can be inferred from Table \ref{tb:1}.} \label{fg:Ratios}
\end{figure}


The model for the intramolecular autocorrelation $G_{R}(t)$ is plotted in Fig. \ref{fg:Gt}(a) as the dashed black lines. The model $G_{R}(t)$ is an exponential decay, which looks like a straight-line decay on the semilog-$y$ plot. It is clear that none of the intramolecular $G_R(t)$ resemble an exponential decay, but appear to be more ``stretched" in nature. 
What is more remarkable from Fig \ref{fg:Gt}(a) is that even the intramolecular autocorrelation $G_R(t)$ for water is {\it not} exponential. $G_R(t)$ for water only involves two $I = 1/2$ spins, and therefore cannot contain cross-correlation effects such as predicted for three or more spins \cite{hubbard:jcp1969}. In the present case, cross-correlation effects originate from interference between different interaction mechanisms, such as a dipole-dipole interaction between two spins and a dipole-dipole interaction between a different pair of spins \cite{frueh:pnmrs2002}. Given that cross-correlation cannot exist for intramolecular $G_R(t)$ in water indicates that the observed disagreement with Eq. \ref{eq:Gmodel} is due to the non-spherical nature of the molecules, and not cross-correlation effects. 

It also should be clarified that a ``stretched" autocorrelation function $G_R(t)$ observed in Fig. \ref{fg:Gt}(a) does {\it not} imply a ``stretched" or multi-exponential relaxation $T_{1,2,R}$. All of the measured $T_2$ \cite{shikhov:amr2016} and $T_1$ \cite{lo:SPE2002} data are consistent with single-exponential decays. The nature of the  ``stretched" correlation function $G_R(t)$ is simply a reflection that the underlying interaction is not well modeled by hard spheres, and that a more sophisticated theory is required.

Similar observations and conclusions can be made of the intermolecular $G_T(t)$ in Fig. \ref{fg:Gt}(b), where the alkanes appear to be more ``stretched" in nature than the model, and increasingly so with larger chain-lengths. $G_T(t)$ for water appears more consistent with the model, but still shows deviations. One informative way to quantify deviations from the hard-sphere model is to plot the ratio ${\tau_D}/{\tau_R}$, which according to Eq. \ref{eq:Gmodel} should be a constant ${\tau_D}/{\tau_R} = 9$ \cite{abragam:book} if the Stokes-Einstein radius for rotation $R_R$ equals the Stokes-Einstein radius for translation $R_T$. The ratio plotted in Fig. \ref{fg:Ratios}(b) clearly shows that lower alkanes tend towards the expectations of the model, while the higher alkanes increasingly depart from the model. This finding again confirms that the lower alkanes are more ``spherical" than the higher alkanes, but surprisingly water, which on the basis of the molecular structure should be considered closest to being ``spherical," has relaxation times similar to octane. This emphasizes that correct account of  intermolecular interactions, crucially short-range hydrogen bonding in the case of water, is necessary in capturing the relaxation behavior. This adds an additional level of complexity to the analysis, but one that is captured by the MD simulations.  More discussions of the Stokes-Einstein radii are given below.

For completeness, the Fourier transform (Eq. \ref{eq:FourierRTcos}) of the autocorrelation functions Eq. \ref{eq:Gmodel} leads to the following spectral densities for the model \cite{cowan:book}:
\begin{eqnarray}
\begin{aligned}
J_R(\omega) &= \frac{2}{3} \Delta\omega_R^2 \tau_R  \frac{1}{1+\left(\omega\tau_R\right)^2},\\
J_T(\omega) &= \frac{2}{3} \Delta\omega_T^2 \tau_T 
\int_{0}^{\infty} \frac{15}{2} \frac{xJ_{3/2}^2(x)}{x^4 + \left(\frac{5}{2}\omega \tau_T\right)^2} dx.
\end{aligned}  	\label{eq:Jmodel}
\end{eqnarray}
Both expressions correctly reduce to $J_{R,T}(0)$ (Eq. \ref{eq:JRTmotional}) in the fast motion regime $\omega_0\tau_R \ll 1$ and $\omega_0\tau_T \ll 1$. The origin of the $\tau_D = \frac{5}{2}\tau_T$ factor comes from normalizing the $J_T(\omega)$ integral in Eq. \ref{eq:Jmodel}, while keeping the 2/3 pre-factor the same as $J_R(\omega)$ \cite{cowan:book}. 

In the hard-sphere model, the intramolecular interaction is based on a rotating hard sphere where the dipole-dipole distances $r_{ij}(t)$ are assumed to be rigid, i.e. $r_{ij}(t) = r_{ij}$ are time independent. This simplifies the expression for the second moment $\Delta\omega_R^2$ in Eq. \ref{eq:SMRTuncorr} to the following:
\begin{eqnarray}
\begin{aligned}
\Delta\omega^2_R &=  \frac{9}{20} \! \left(\frac{\mu_0}{4\pi}\right)^{\! 2} \! \hbar^2 \gamma^4 \frac{1}{N_R}\sum\limits_{i \neq j}^{N_R}{\frac{1}{r_{ij}^6}}, \\
\Delta\omega^2_T &= \frac{3\pi}{40} \! \left(\frac{\mu_0}{4\pi}\right)^{\! 2} \! \hbar^2 \gamma^4 \frac{N_V}{R_T^3}, \\
\label{eq:SMmodel}
\end{aligned} 
\end{eqnarray}
where $\Delta\omega_R^2$ is consistent with previous definitions \cite{foley:JMR1996}. The full MD simulation for $\Delta\omega_R^2$ with non-rigid bonds (Eq. \ref{eq:SMRTuncorr}) are slightly higher $\lesssim 1 \%$ than with rigid bonds (Eq. \ref{eq:SMmodel}). 

The hard-sphere model for the intermolecular interaction is based on diffusing hard spheres with a hard-core distance $2R_T$. The model yields the above expression for $\Delta\omega_T^2$ in terms of $R_T$ and the density of $^1$H spins defined as $N_V$. The following expression is used $N_V = \rho \ n_H N_A/M$, where $n_H$ is the number of $^1$H per formula unit, $N_A$ is Avogadro's number, $M$ is the molar mass, and $\rho$ is the density from Table \ref{tb:2}. 

\subsection{Molecular Radius}\label{sc:Radius}

\begin{figure}
	\begin{center}
		\includegraphics[width=0.9\columnwidth]{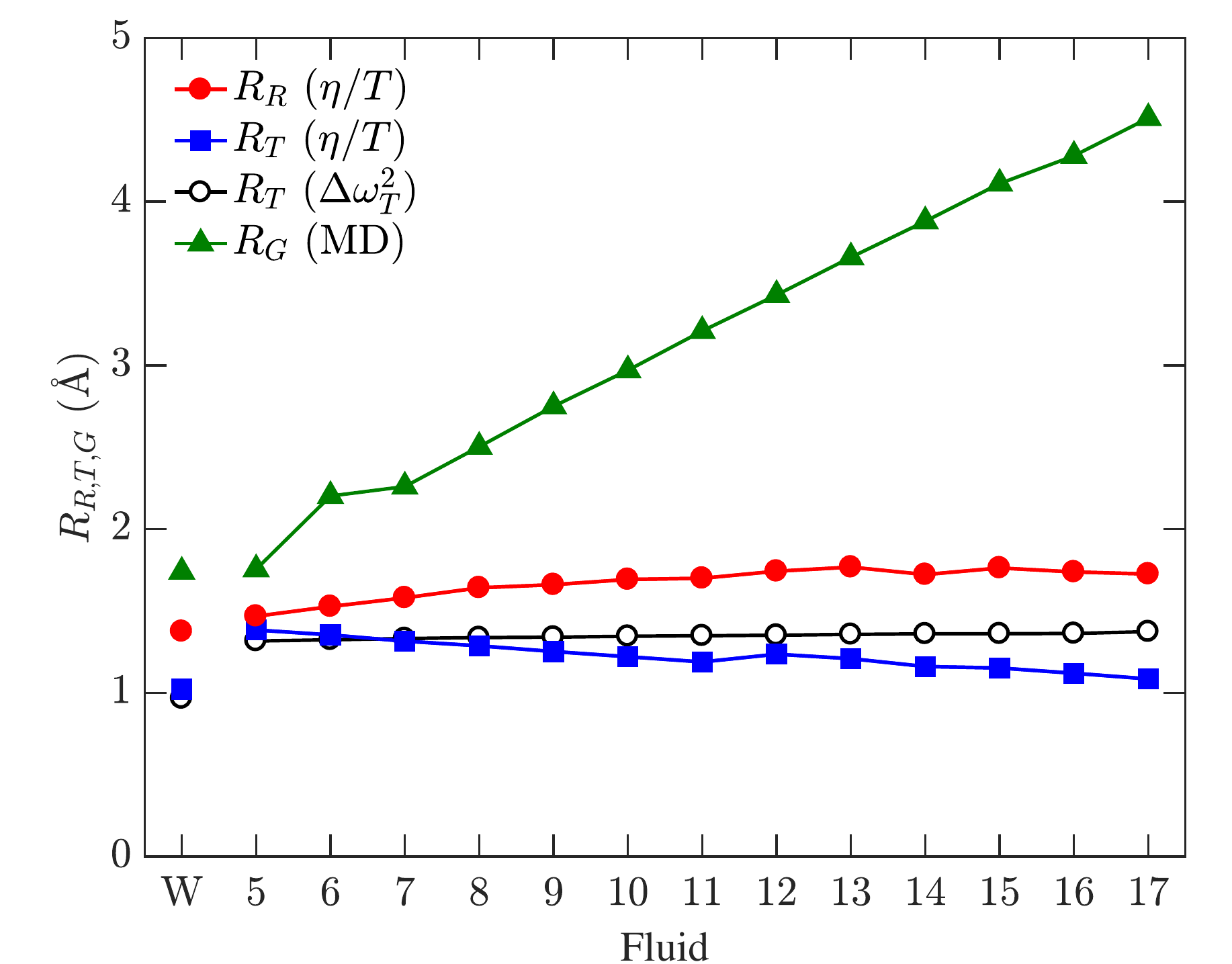} 		
	\end{center}
	\caption{Stokes-Einstein radius for rotational motion $R_R$ from Eq. \ref{eq:TauModel} (using $\eta/T$, with $\eta$ listed in Table \ref{tb:2})), Stokes-Einstein radius for translational motion $R_T$ from Eq. \ref{eq:TauModel} (using $\eta/T$), $R_T$ prediction from second moment $\Delta\omega_T^2$ from Eq. \ref{eq:SMmodel} (using $\rho$ from Table \ref{tb:1}), and $R_G$ from MD simulations (Eq. \ref{eq:Rg}), for water (W) and alkanes (carbon number C5$\rightarrow$C17). Data are at $20^{\rm o}$C, and are listed in Table \ref{tb:2}.} \label{fg:Radius}
\end{figure}

It is informative to compare the various sets of molecular radii that result from MD simulations and the Stokes-Einstein relation in Eq. \ref{eq:TauModel}. As shown in Fig. \ref{fg:Ratios}(b), the predicted ratio of translational diffusion to rotational correlation times are $\tau_D/\tau_R < 9 $, which implies that $R_R > R_T$. Using Eq. \ref{eq:TauModel} and viscosity $\eta$ data in Table \ref{tb:1} results in the Stokes-Einstein radii $R_R$ and $R_T$, which are listed in Table \ref{tb:2} and plotted in Fig. \ref{fg:Radius}. While $R_R \simeq R_T$ in the case of pentane, the rotational radius $R_R$ increases mildly with carbon number, while $R_T$ mildly {\it decreases} with increasing carbon number. These trends indicate that the molecules become less ``spherical" with increasing chain length.

\begin{table}
	\begin{center}
		\begin{tabularx}{8.5cm}{CCCCCCCC}		
\hline

		\multicolumn{2}{c}{$^{{\strut}{}}$Fluid} & $\eta$ & $\rho$ & $R_R$ & $R_T$ & $R_T$  & $R_G$  \\ 
		$^{{\strut}{}}$& & (cP)& (g/cm$^3$)& $(\rm{\AA})$ & $(\rm{\AA})$ & $(\rm{\AA})$ & $(\rm{\AA})$ \\ 
		$^{{\strut}{}}$& & Ref. \cite{viscosity}& Ref. \cite{nist,*crc,viscosity}& Eq. \ref{eq:TauModel} & Eq. \ref{eq:TauModel}  & Eq. \ref{eq:SMmodel} &  MD \\ \hline

\multicolumn{2}{c}{$^{{\strut}{}}$H$_2$O}	&	1.00	&	0.998	&	1.38	&	1.02	&	0.97	&	1.74	\\
\multicolumn{2}{c}{$n^{{\strut}{}}$-C$_{5}$H$_{12}$}	&	0.24	&	0.626	&	1.47	&	1.38	&	1.32	&	1.76	\\
\multicolumn{2}{c}{$n^{{\strut}{}}$-C$_{6}$H$_{14}$}	&	0.31	&	0.659	&	1.53	&	1.35	&	1.32	&	2.20	\\
\multicolumn{2}{c}{$n^{{\strut}{}}$-C$_{7}$H$_{16}$}	&	0.42	&	0.684	&	1.58	&	1.32	&	1.33	&	2.26	\\
\multicolumn{2}{c}{$n^{{\strut}{}}$-C$_{8}$H$_{18}$}	&	0.55	&	0.702	&	1.64	&	1.29	&	1.34	&	2.50	\\
\multicolumn{2}{c}{$n^{{\strut}{}}$-C$_{9}$H$_{20}$}	&	0.72	&	0.718	&	1.66	&	1.25	&	1.34	&	2.75	\\
\multicolumn{2}{c}{$n^{{\strut}{}}$-C$_{10}$H$_{22}$}	&	0.93	&	0.731	&	1.69	&	1.22	&	1.35	&	2.97	\\
\multicolumn{2}{c}{$n^{{\strut}{}}$-C$_{11}$H$_{24}$}	&	1.19	&	0.740	&	1.70	&	1.19	&	1.35	&	3.21	\\
\multicolumn{2}{c}{$n^{{\strut}{}}$-C$_{12}$H$_{26}$}	&	1.51	&	0.750	&	1.74	&	1.24	&	1.35	&	3.43	\\
\multicolumn{2}{c}{$n^{{\strut}{}}$-C$_{13}$H$_{28}$}	&	1.89	&	0.756	&	1.77	&	1.21	&	1.36	&	3.66	\\
\multicolumn{2}{c}{$n^{{\strut}{}}$-C$_{14}$H$_{30}$}	&	2.34	&	0.760	&	1.72	&	1.16	&	1.36	&	3.88	\\
\multicolumn{2}{c}{$n^{{\strut}{}}$-C$_{15}$H$_{32}$}	&	2.87	&	0.769	&	1.76	&	1.15	&	1.36	&	4.11	\\
\multicolumn{2}{c}{$n^{{\strut}{}}$-C$_{16}$H$_{34}$}	&	3.48	&	0.773	&	1.74	&	1.12	&	1.36	&	4.28	\\
\multicolumn{2}{c}{$n^{{\strut}{}}$-C$_{17}$H$_{36}$}	&	4.21	&	0.778	&	1.73	&	1.09	&	1.37	&	4.51	\\

\hline
		\end{tabularx}
	\end{center}
	\caption{Molecular radii, including: Stokes-Einstein radius for rotational motion $R_R$ from Eq. \ref{eq:TauModel} (using $\eta/T$, with listed $\eta$ taken from \cite{viscosity}), Stokes-Einstein radius for translational motion $R_T$ from Eq. \ref{eq:TauModel} (using $\eta/T$), $R_T$ prediction from second moment $\Delta\omega_T^2$ from Eq. \ref{eq:SMmodel}, and $R_G$ directly from MD simulations (Eq. \ref{eq:Rg}). Sources for density data \cite{nist,*crc} include the NIST chemistry webbook for $n$-C$_{5}$H$_{12}$ to $n$-C$_{10}$H$_{22}$, CRC Handbook for $n$-C$_{11}$H$_{24}$ to $n$-C$_{5}$H$_{12}$, and API tech report \cite{viscosity} for $n$-C$_{15}$H$_{32}$ to $n$-C$_{17}$H$_{36}$. All data are at 20$^{\rm 0}$C.
		\label{tb:2} }
\end{table}

Another prediction from the hard-sphere model is $R_T$ according to $\Delta\omega^2_T$ defined in Eq. \ref{eq:SMmodel}. Using the values of $N_V$ and $\Delta\omega^2_T$ in Table \ref{tb:1} results in a prediction for $R_T$, which is listed in Table \ref{tb:2} and plotted in Fig. \ref{fg:Radius}. Since both $N_V$ and $\Delta\omega^2_T$ show little variation with carbon number, the resulting $R_T$ is also found to be roughly independent of carbon number. In the case of pentane, a consistent value of $R_T$ according to $\eta/T$ and $\Delta\omega^2_T$ confirm that the model is more accurate for more ``spherical" molecules. 

The radius of gyration of the molecule, defined by
\begin{equation}
R_G = \sqrt{\left<\frac{1}{N}\sum\limits_{i=1}^{N}\left({\bf r}_i - {\bf r}_{{\rm cm}}\right)^2 \right>_{\!\! \tau}}
\label{eq:Rg}
\end{equation}
provides an independent estimate of the size of the molecule; for a spherical molecule the $R_G$ is expected to be a reasonable estimate of the hydrodynamic radius. Here $R_G$ is calculated by sampling a single molecule from the simulation and averaging over the entire trajectory. As expected, increasing the sample size of the simulation did not change the results. The simulation results for $R_G$ are listed in Table \ref{tb:2} and plotted in Fig. \ref{fg:Radius}. 

The results in Fig. \ref{fg:Radius} clearly show that while $R_G$ increases linearly with chain length, both $R_R$ and $R_T$ show only mild variations with increasing chain length. This suggests that the NMR related quantities $R_R$ and $R_T$ remain more local in nature than the 
structure-based $R_G$ radius. In the case of water, the pattern $R_G > R_R > R_T$ again emphasizes that it is best to not regard water as a spherical object, as required by the traditional hard-sphere model. 

\section{Conclusions}\label{sc:Conc}

MD simulation techniques for predicting NMR relaxation and diffusion in bulk hydrocarbons and water are validated against NMR measurements, without any adjustable parameters in the interpretation of the simulation data. MD simulations reveal new insight about the intramolecular versus intermolecular NMR relaxation in bulk fluids, which were not easily accessible before. The simulations quantify the relative strength of the two relaxation mechanisms, indicating that intramolecular increasingly dominates over intermolecular relaxation with increasing molecular chain-length (i.e. increasing carbon number). This validates the common practice of only considering intramolecular dipole-dipole interactions for simulations of macromolecules such as proteins and polymers. 

With increasing chain length, the simulations indicate increasing departure from the traditional hard-sphere models for the time dependence in autocorrelation functions of both intramolecular and intermolecular dipole-dipole interactions. The MD simulations indicate that the Stokes-Einstein radius for rotational and translational motion are comparable for small molecules, but increasingly diverge with increasing chain length, indicating a departure from ``spherical" molecules for the higher alkanes. Furthermore, with increasing chain length the radius of gyration is found to increase more rapidly than the NMR derived Stokes-Einstein radius, indicative of the more local nature of the NMR measurement. 


The validation of MD simulations for predicting NMR relaxation and diffusion of bulk hydrocarbons and water opens up new opportunities for investigating more complex scenarios, such as the effects of nanometer confinement on the phase-behavior and transport properties of hydrocarbons and water in organic porous-media such as kerogen. It also opens up opportunities for studying NMR relaxation in complex fluids such as crude-oils or model systems such as polymer-solvent mixes. 

\section*{Acknowledgments}\label{sc:Acknow}
We gratefully acknowledge funding from the Rice University Consortium on Processes in Porous Media.
We gratefully acknowledge the National Energy Research Scientific Computing Center, which is supported by the Office of Science of the U.S. Department of Energy [DE-AC02-05CH11231], for HPC time and support. We also gratefully acknowledge the Texas Advanced Computing Center (TACC) at The University of Texas at Austin (URL: http://www.tacc.utexas.edu) for providing HPC resources; Arjun Valiya~Parambathu, Jinlu Liu, and Zeliang Chen for their assistance; and, Igor Shikhov and Christoph H. Arns for sharing their experimental NMR data.


%

\end{document}